# Mass and environment as drivers of galaxy evolution in SDSS and zCOSMOS and the origin of the Schechter function[1]

Ying-jie Peng[2], Simon J. Lilly[2], Katarina Kovač[2], Micol Bolzonella[3], Lucia Pozzetti[3], Alvio Renzini[4], Gianni Zamorani[3], Olivier Ilbert[5], Christian Knobel[2], Angela Iovino[6], Christian Maier[2], Olga Cucciati[5], Lidia Tasca[7], C. Marcella Carollo[2], John Silverman[2], Pawel Kampczyk[2], Loic de Ravel[5], David Sanders[8], Nicholas Scoville[9], Thierry Contini[10], Vincenzo Mainieri[11], Marco Scodeggio[7], Jean-Paul Kneib[5], Olivier Le Fèvre[5], Sandro Bardelli[3], Angela Bongiorno[12], Karina Caputi[2], Graziano Coppa[3], Sylvain de la Torre[5], Paolo Franzetti[7], Bianca Garilli[7], Fabrice Lamareille[10], Jean-Francois Le Borgne[10], Vincent Le Brun[5], Marco Mignoli[3], Enrique Perez Montero[10], Roser Pello[10], Elena Ricciardelli[3], Masayuki Tanaka[12], Laurence Tresse[5], Daniela Vergani[3], Niraj Welikala[5], Elena Zucca[3], Pascal Oesch[2], Ummi Abbas[5], Luke Barnes[2], Rongmon Bordoloi[2], Dario Bottini[7], Alberto Cappi[3], Paolo Cassata[5], Andrea Cimatti[13], Marco Fumana[7], Gunther Hasinger[12], Anton Koekemoer[14], Alexei Leauthaud[15], Dario Maccagni[7], Christian Marinoni[16], Henry McCracken[18], Pierdomenico Memeo[7], Baptiste Meneux[11], Preethi Nair[3], Cristiano Porciani[2], Valentina Presotto[6], Roberto Scaramella[17]

*Footnotes and Author Addresses*

1   Based on observations undertaken at the European Southern Observatory (ESO) Very Large Telescope (VLT) under Large Program 175.A-0839. Also based on observations with the NASA/ESA Hubble Space Telescope, obtained at the Space Telescope Science Institute, operated by AURA Inc., under NASA contract NAS 5-26555, with the Subaru Telescope, operated by the National Astronomical Observatory of Japan, with the telescopes of the National Optical Astronomy Observatory, operated by the Association of Universities for Research in Astronomy, Inc. (AURA) under cooperative agreement with the National Science Foundation, and with the Canada-France-Hawaii Telescope, operated by the National Research Council of Canada, the Centre National de la Recherche Scientifique de France and the University of Hawaii.
2.  Institute for Astronomy, ETH Zurich, Wolfgang-Pauli-strasse 27, 8093 Zurich, Switzerland
3.  INAF Osservatorio Astronomico di Bologna, Via Ranzani 1, 40127 Bologna, Italy
4.  INAF - Osservatorio Astronomico di Padova, Vicolo dell'Osservatorio 5, 35122, Padova, Italy
5.  Laboratoire d'Astrophysique de Marseille, 38, rue Frédéric Joliot-Curie, 13388 Marseille Cedex 13, France
6.  INAF Osservatorio Astronomico di Brera, via Brera 28, 20121 Milano
7.  INAF - IASF Milano, Via E. Bassini 15, 20133 Milano, Italy
8.  University of Hawaii, 2680 Woodlawn Drive, Honolulu, Hawaii 96822-1839, USA
9.  Department of Astronomy, Caltech, MC 105-24, 1200 East California Blvd, Pasadena CA 91125, USA
10. Laboratoire d'Astrophysique de Toulouse/Tarbes, Université de Toulouse, 14 avenue Edouard Belin, 31400 Toulouse, France
11. European Southern Observatory, Karl-Schwarzschild-Strasse 2, 85748 Garching, Germany
12. Max-Planck-Institut für extraterrestrische Physik, Giessenbachstraße, 85748 Garching, Germany.
13. Dipartimento di Astronomia, Universita` degli Studi di Bologna, Via Ranzani 1, 40127 Bologna, Italy
14. Space Telescope Science Institute, 3700 San Martin Drive, Baltimore, Maryland 21218, USA.
15. Berkeley Lab & Berkeley Center for Cosmological Physics, University of California, Lawrence Berkeley National Lab, 1 Cyclotron Road, MS 50-5005, Berkeley, CA 94720, USA.
16. Centre de Physique Theorique, Case 907 - 13288 Marseille cedex 9, France
17. INAF - Osservatorio Astronomico di Roma, Osservatorio Astronomico di Roma, Via di Frascati 33, 00040 Monte Porzio Catone, Italy
18. Institut d'Astrophysique de Paris, UMR7095 CNRS, Université Pierre & Marie Curie, 75014 Paris, France

ABSTRACT

We explore the simple inter-relationships between mass, star formation rate, and environment in the SDSS, zCOSMOS, and other deep surveys. We take a purely empirical approach in identifying those features of galaxy evolution that are demanded by the data and then explore the analytic consequences of these. We show that the differential effects of mass and environment are completely separable to $z \sim 1$, leading to the idea of two distinct processes of "mass quenching" and "environment quenching." The effect of environment quenching, at fixed over-density, evidently does not change with epoch to $z \sim 1$ in zCOSMOS, suggesting that the environment quenching occurs as large-scale structure develops in the universe, probably through the cessation of star formation in 30%–70% of satellite galaxies. In contrast, mass quenching appears to be a more dynamic process, governed by a quenching rate. We show that the observed constancy of the Schechter $M^*$ and $\alpha_s$ for star-forming galaxies demands that the quenching of galaxies around and above $M^*$ must follow a rate that is statistically proportional to their star formation rates (or closely mimic such a dependence). We then postulate that this simple mass-quenching law in fact holds over a much broader range of stellar mass (2 dex) and cosmic time. We show that the combination of these two quenching processes, plus some additional quenching due to merging naturally produces (1) a quasi-static single Schechter mass function for star-forming galaxies with an exponential cutoff at a value $M^*$ that is set uniquely by the constant of proportionality between the star formation and mass quenching rates and (2) a double Schechter function for passive galaxies with two components. The dominant component (at high masses) is produced by mass quenching and has exactly the same $M^*$ as the star-forming galaxies but a faint end slope that differs by $\Delta\alpha_s \sim 1$. The other component is produced by environment effects and has the same $M^*$ and $\alpha_s$ as the star-forming galaxies but an amplitude that is strongly dependent on environment. Subsequent merging of quenched galaxies will modify these predictions somewhat in the denser environments, mildly increasing $M^*$ and making $\alpha_s$ slightly more negative. All of these detailed quantitative inter-relationships between the Schechter parameters of the star-forming and passive galaxies, across a broad range of environments, are indeed seen to high accuracy in the SDSS, lending strong support to our simple empirically based model. We find that the amount of post-quenching "dry merging" that could have occurred is quite constrained. Our model gives a prediction for the mass function of the population of transitory objects that are in the process of being quenched. Our simple empirical laws for the cessation of star formation in galaxies also naturally produce the "anti-hierarchical" run of mean age with mass for passive galaxies, as well as the qualitative variation of formation timescale indicated by the relative $\alpha$-element abundances.

*Key words:* cosmology: observations – galaxies: active – galaxies: distances and redshifts – galaxies: evolution

## 1. INTRODUCTION

The last few years have seen a flood of new observational data on large samples of galaxies, both locally, as in the Sloan Digital Sky Survey (SDSS; York et al. 2000) and 2dfGRS (Colless et al. 2001), and in large photometric and spectroscopic surveys at higher redshifts, such as COMBO-17 (Wolf et al. 2003), GOODS (Giavalisco et al. 2004), DEEP (Vogt et al. 2005; Weiner et al. 2005), DEEP2 (Davis et al. 2003), VVDS (Le Fèvre et al. 2005), and COSMOS and zCOSMOS (Scoville et al. 2007; Lilly et al. 2007). These new surveys allow an increasingly sophisticated statistical study of the overall properties of the population of galaxies and its evolution over cosmic time.

There has been much work also on developing a theory of galaxy evolution, mostly in the context of so-called semi-analytic models (SAMs) for the galaxy population (e.g., Baugh 2006 for a review), which combine $N$-body simulations of the formation and evolution of dark matter haloes with simple analytic descriptions of all the relevant baryonic physics that can be imagined, including the heating and cooling of gas, the formation of stars, and the merging of galaxies. SAMs have been complemented by increasingly sophisticated hydro-dynamical simulations (e.g., Birnboim & Dekel 2003).

The philosophy of this paper is to take a purely empirical, observation-based approach to the evolving galaxy population. In particular, it is likely that galactic mass and environment are both playing a major role in the evolution of galaxies. Accordingly, we try to identify the most important relations between galaxy properties and their stellar masses and environments in the present-day galaxy population, and in the population at much earlier cosmic times. The goal is to use the observational material as directly as possible in order to identify the simplest things that are apparently *demanded* by the data and thereby to define empirically based "laws" for the evolution of the population.

By identifying and isolating the key underlying trends within different data sets, and then combining them into a simple analytic model for the overall population, we can avoid any difficulties that may be encountered when comparing different observational surveys directly. These may include color transformations at different redshifts or different approaches to the computation of the masses of stellar populations.

We may then try to associate these clear evolutionary signatures with a dominant physical process, but the causal connection cannot of course be proven, and it is quite possible that some different set of physical processes may conspire to mimic the same observed results. Nevertheless, our identification of the most important empirical characteristics of the evolution serves to constrain the permitted outcomes of the physical processes involved and may help to illuminate the most important parameters that apparently control galaxy evolution. This approach may be regarded as a kind of "purely empirical analytic model" for galaxy evolution.

In this paper, we focus primarily on the processes that evidently cause the cessation of star formation in some star forming galaxies and lead to the emergence of the so-called "red-sequence" of passive galaxies. We refer to this cessation of star formation as "quenching," regardless of its physical cause, and whether it is internally or externally induced.

Quenching is thus distinct from the general decline in the specific star-formation rate of *star-forming* galaxies that has occurred between $z \sim 2$ and the present, whose cause is not well understood but which may be linked to the dwindling supply of gas onto galaxies. Quenching in contrast is assumed to produce passive galaxies in which the star-formation rate is very low, or zero, leading to the familiar bi-modality in the galaxy population.

Our primary goal is to understand how the quenching of galaxies depends on galaxy stellar mass (henceforth $m$), on environment (henceforth the density $\rho$ or over-density $\delta$), and on the cosmic epoch, $t$. However, in order to understand the effects of quenching, we must also consider the growth in stellar mass of non-quenched star-forming galaxies through star-formation. We therefore emphasize also the simple relations between *star-formation rates* and stellar mass, environment and epoch.

Our analysis is built on the following three key observational facts about the galaxy population that we take from the literature, or establish in the current paper:

1. The differential effects of galactic mass and of the environment in the quenching of galaxies are fully separable. This new result is shown to hold both in the SDSS (Section 4.2) and zCOSMOS (Section 4.3) at least out to $z \sim 1$, and we will assume that it applies also at higher redshifts.
2. The Schechter mass-function of *star-forming* galaxies is remarkably constant in terms of M* and $\alpha_s$ and, to a lesser degree, in $\phi$* (Bell et al., 2003, 2007, Pozzetti et al., 2009, Ilbert et al., 2010). This has been now clearly demonstrated to $z \sim 2$.
3. The specific star-formation rate sSFR($m,t$) of star-forming galaxies is at most a weak function of galactic stellar mass, and falls sharply between $z = 2$ (Daddi et al., 2007, Elbaz et al., 2007, Noeske et al., 2007) and the present. We show in Section 3.2 that this sSFR($m,t$) is also evidently independent of environment out to $z \sim 1$, and we will assume that this is also true at higher redshifts. The simple behaviour of the sSFR with mass and environment greatly simplifies our analysis, but is not strictly required for the validity of most of the conclusions.

We additionally take empirical estimates of the merging rate of galaxies from our own zCOSMOS analyses (de Ravel et al., Kampczyk et al., in preparation) and from the literature when required.

The separability of the effects of mass and environment suggests that there are two independent processes operating. The above observational facts allow us to identify two striking observational signatures associated with each of these processes which, together with an observationally determined merging rate, successfully account for some of the most basic features of the galaxy population, most notably the inter-relationships between the parameters describing the mass-functions of star-forming and passive galaxies.

The layout of the paper is as follows. In Section 2, we review the basic input data that we have used from SDSS and zCOSMOS. This includes the derivation of a new density field for the SDSS sample that is consistent with our zCOSMOS density field.

In Section 3, we briefly review the behaviour of sSFR($m,t$), and show that in SDSS, and also in zCOSMOS, the sSFR of star-forming galaxies is independent of environment, even though the fraction of galaxies that are star-forming can depend quite strongly on environment. We also review the measurements of the mass-function of star-forming galaxies back to $z \sim 2$.

In Section 4, we introduce a new formalism to examine the *differential* effects of mass and Mpc-scale environment on the fraction of galaxies that have been quenched, $f_{red}(m, \rho)$, at a given mass and in a given environment. We demonstrate that the effects of mass and environment are fully separable in the SDSS sample, indicating that two distinct processes are occurring, which we henceforth refer to as "mass quenching" and "environment quenching". We then look at how this scheme evolves in the zCOSMOS data out to $z \sim 1$. This leads us to identify a clear signature of the environment quenching process.

The effects of mass quenching are however more clearly seen from consideration of the mass-function of star-forming galaxies, which reflects the population of "surviving" galaxies, rather than from the red fraction which mixes the living and the dead. Here too, we show that the available data demand a particularly simple "law" for the mass-quenching process.

We argue in Section 5 that these two remarkably simple and empirically defined processes appear to control many of the gross features of the galaxy population. In particular, our very simple empirically-based model naturally:

1. establishes a pure Schechter mass-function for star forming galaxies, and sets the characteristic mass M*;
2. produces a two-component Schechter mass-function for passive galaxies, and for all galaxies (active plus passive) combined, and predicts well-defined relationships between the Schechter parameters of the various components that are observed in the galaxy population, with only small modifications due to some limited subsequent merging of galaxies;
3. accounts qualitatively for several other simple observational features of the galaxy population, such as the mean age-mass relation for passive galaxies and the $\alpha$-enrichment of the more massive passive galaxies (presented in Section 7).

In Section 6 we construct a simple simulation of the evolving galaxy population based on the remarkably simple picture outlined above, i.e. on just 3-4 observationally determined parameters, and show that from an initial starting point at $z \sim 10$ this successfully reproduces the mass-function and $f_{red}$ of the SDSS sample as a function of environment.

Not surprisingly, it is possible to associate these two strikingly simple evolutionary signatures with two of the main physical processes that have been introduced into semi-analytic models of galaxy formation, namely satellite quenching as galaxies fall into larger dark matter haloes (our environment quenching), and feedback processes (our mass quenching). We believe that their remarkably simple action is very clearly demonstrated in the current purely empirical analysis, which serves to highlight those simple signatures of these processes that must be understood from a more physically-based standpoint. As noted above, it is of course possible that other combinations of physical processes may mimic these observational signatures.

The cosmological model used in this analysis is a concordance LCDM cosmology with $H_0 = 70$ kms$^{-1}$Mpc$^{-1}$, $\Omega_\Lambda = 0.75$ and $\Omega_M = 0.25$. All magnitudes are quoted in the AB normalization. Throughout the paper we use the term "dex" to mean the anti-logarithm, i.e. 0.1 dex = $10^{0.1} = 1.258$.

## 2. OBSERVATIONAL DATA

### 2.1 zCOSMOS

#### 2.1.1 Sample

The zCOSMOS spectroscopic survey (Lilly et al. 2007) is a large redshift survey that is being undertaken in the ~1.7 deg$^2$ COSMOS field (Scoville et al., 2007). The survey is designed to characterize the environments of COSMOS galaxies from the 100 kpc scales of galaxy groups up to the 100 Mpc scale of the cosmic web and to produce diagnostic information on galaxies and active galactic nuclei (AGN). It is thus ideally suited to the study of the effects of environment in galaxy evolution. The zCOSMOS-bright survey will eventually contain over 20,000 galaxies, with a pure flux limited selection at $I_{AB} < 22.5$, yielding $0.1 < z < 1.4$. The zCOSMOS-deep component will contain several thousand galaxies at higher redshifts $1.4 < z < 3$. The results described in this paper are based on the first 10,644 redshifts measured in the bright sample (Lilly et al. 2009), which is hereafter called the "10k-sample". The reader is referred to a number of detailed studies of the properties of galaxies in different environments that have been undertaken using this sample (Cucciati et al., 2009, Tasca et al., 2009, Iovino et al., 2009, Kovač et al., 2009, Vergani et al., 2010, Zucca et al., 2009).

#### 2.1.2 Star-formation rates and masses

Rest-frame colors and absolute magnitudes for the zCOSMOS 10k sample are derived from the spectral energy distributions (SEDs) obtained by applying the ZEBRA photo-z code (Feldmann et al. 2006) to the best available COSMOS photometry (Ilbert et al., 2010) after application of small zero-point offsets. The adopted SED for each galaxy is that of the best-fit template at the spectroscopic redshift.

Galaxy stellar masses, hereafter indicated by $m$, are computed as in Bolzonella et al. (2009), to which the reader is referred for details. Briefly, these are based on the *Hyperzmass* code (Bolzonella et al., 2009) with a set of ten exponentially decreasing star-formation histories with *e*-folding time scales $\tau$ ranging from 0.1 to 30 Gyr, plus one model with constant star formation. We adopted a Calzetti et al. (2000) extinction law, solar metallicity, and Bruzual & Charlot (2003) population synthesis models with a Chabrier initial mass function (Chabrier et al. 2003) with lower and upper cutoffs of 0.1 and 100 $M_\odot$. Galaxy stellar masses are calculated by integrating the star formation rate over the galaxy age and subtracting the "return fraction", which is the mass of gas processed by stars and returned to the interstellar medium during their evolution.

Where necessary, star-formation rates are taken as the instantaneous star-formation rate in the best-fitting model. We have compared these star-formation rates with those implied by the strength of the [OII] $\lambda$ 3727 emission line in the redshift range $0.5 < z < 0.9$ where the line is accessible (Maier et al. 2009). In general, a good correlation between the two star formation rates is found (see discussion in Pozzetti et al., 2009).

#### 2.1.3 Density field

We use the zCOSMOS-10k density field constructed by Kovač et al. (2010a, hereafter K10a). This is based on application of the ZADE algorithm, which combines known spectroscopic redshifts with the photo-z of objects not yet observed spectroscopically. This is done by modifying the photo-z redshift probability distribution functions using the spectroscopic redshifts of nearby galaxies along the line of sight. The reader is referred to K10a for a full description of the method and the extensive tests that have been undertaken of its performance and reliability. The environment of a given zCOSMOS galaxy is characterized by the dimensionless density contrast, or over-density, $\delta_i$, defined as $\delta_i = (\rho_i - \rho_m)/\rho_m$, where $\rho_m$ is the (volume) mean density at a given redshift. We choose the "unity-weighted" density (counting un-weighted galaxies) with the "5th nearest neighbor" density estimator (hereafter 5NN) of K10a.

It should be noted that the densities used here are locally projected densities, in that both the calculation of the distance to the fifth nearest neighbor, and the computation of the resulting densities, are undertaken over a cylindrical volume of length (in the radial dimension) corresponding to ±1000 km s$^{-1}$. In this paper, we use the density fields defined by two approximately "volume-limited" samples of tracers: the fainter one is defined by $M_{B,AB} \leq -19.3 - z$, which is accessible in zCOSMOS-bright for all $z \leq 0.7$. As in K10a, we adopt at higher redshifts $0.7 < z \leq 1.0$ a brighter tracer sample with $M_{B,AB} \leq -20.5 - z$. The $-z$ term in the above limiting magnitudes accounts, at least approximately, for the luminosity evolution of individual galaxies. Adopting these tracer populations gives a density field that samples the underlying density field (at least as traced by these galaxies) on scales that are about one comoving Mpc for typical galaxies (see K10a for details).

#### 2.1.4 Treatment of spectroscopic- and mass-incompleteness

The zCOSMOS-bright survey is overall only about 90% complete in successful redshift determination, although this increases to about 95% for $0.5 < z < 0.8$ (Lilly et al., 2009). Therefore, statistical weights were applied to all objects with secure spectroscopic redshifts in constructing the population statistics. Each galaxy was weighted by $1/TSR \times 1/SSR$, where *TSR* is the spatial Target Sampling Rate, easily derived from the spatial distribution of target and spectroscopically observed objects, and *SSR* is a Spectroscopic Success Rate, constructed using the photo-z of objects for which we failed to obtain a spectroscopic redshift (see Bolzonella et al. 2009 and Zucca et al. 2009 for further details). The TSR is included to account for any residual correlation between environmental richness and sampling rate (although this is negligible). We also apply a $1/V_{max}$ weighting to account for any residual volume incompleteness within a given redshift bin, using $V_{max}$ values derived from *k*-correcting the ZEBRA SED fits.

Finally, it should be noted that zCOSMOS-bright sample is a luminosity-selected sample and that the stellar mass completeness of the sample therefore depends quite strongly on both the redshifts and the range of mass-to-light ratios at the survey limit, i.e. on the galaxy SEDs. We use the $M_{bias}$, the lowest mass at which the sample can be considered to be complete at a given redshift, as constructed in Pozzetti et al. (2009). In the current work, we consider only mass-complete sub-samples at $m > M_{bias}$.

## 2.2 Sloan Digital Sky Survey (SDSS)

### 2.2.1 Construction of the sample

The local comparison sample is based on the SDSS seventh data release (DR7) (Abazajian et al. 2009), which is the final public version. Our parent SDSS sample was retrieved directly from the SDSS CasJobs site. Following Baldry et al. (2006), we first select galaxies in "Galaxy View" which have clean photometry, Petrosian $r$ magnitudes in the range of $10.0 < r < 18.0$ after correction for Milky-Way galactic extinction, and $r_{PSF} - r_{Model} > 0.25$ to exclude stars. We then use 'SpecObj View' to select objects with clean spectra. This produced the parent sample of 1,579,314 objects after removing duplicates, of which 238,474 objects have reliable spectroscopic redshifts measurements in the redshift range $0.02 < z < 0.085$. These comprise the SDSS sample used henceforth in this paper.

Due to the minimum fiber spacing of 55 arcsec, about 10% of the SDSS targets are missed from the spectroscopy sample. To correct for this, a *TSR* was determined using the fraction of objects that have spectra in the parent photometric sample within 55 arcsec of a given object. In constructing the population of SDSS galaxies, and in computing the density field, galaxies are weighted by 1/*TSR* to account for any linkage between the sampling rate and the local environment, and hence other properties, of a galaxy.

The SDSS spectroscopic selection $r < 17.77$ is only complete at $z = 0.085$ above a stellar mass of about $10^{10.4} M_\odot$. Because we wish to consider the population of galaxies at lower masses in our analysis, we weight galaxies below this stellar mass limit using the $V_{max}$ method, employing the $V_{max}$ values from the *k-correction* program v4_1_4 (Blanton et al. 2007). In constructing the final "population" of SDSS galaxies, we therefore weight each galaxy by $1/TSR \times 1/V_{max}$ where $V_{max} = 1$ for galaxies above this mass limit.

### 2.2.2 Star-formation rates and masses

Rest-frame absolute magnitudes for the SDSS sample are derived from the five SDSS *ugriz* bands using the *k-correction* program (Blanton et al. 2007). All SDSS magnitudes are further corrected onto the AB magnitude system. To check for consistency between our SDSS and zCOSMOS derived colors we have computed the *ugriz*-based (U-B) colors for roughly 200 low redshift objects with $r < 19.3$ for which we have zCOSMOS redshifts, and find negligible systematic offset. The stellar masses are determined directly from the same *k-correction* code with Bruzual & Charlot (2003) population synthesis models and a Chabrier IMF. The derived stellar masses were then compared with the published stellar masses of Kauffmann et al. (2003a) and Gallazzi et al. (2005). They show an encouragingly small scatter of about 0.1 dex. Comparison of the derived masses for the overlap objects also shows negligible offsets between SDSS and zCOSMOS. We further tested the masses with the new version of the Charlot & Bruzual (in preparation) library.

The SFR for the SDSS blue star-forming galaxies was taken from Brinchmann et al. (2004, hereafter B04). These are based on the Hα emission line luminosities, corrected for extinction using the Hα/Hβ ratio and corrected for aperture effects. The B04 SFR was computed for a Kroupa IMF and so we convert these to a Chabrier IMF, by using log *SFR* (Chabrier) = log *SFR* (Kroupa) - 0.04.

### 2.2.3 Construction of the density field

We have computed a comoving density $\rho$ and an over-density $\delta$ for all galaxies in the SDSS sample in as similar a way as we can to the zCOSMOS approach that we described above. We use the same volume-limited tracer population of $M_{B,AB} \leq -19.3 - z$, and compute the "unity-weighted" 5NN density field over the redshift range $0.02 \leq z \leq 0.085$, checking that there is little difference with the density field that would be obtained using the stronger evolution -1.6$z$ preferred by Blanton et al. (2003). We again use projected densities in cylinders corresponding to an interval of ±1000 kms$^{-1}$. Since the effect of incomplete spatial sampling is small (only ~10% of the SDSS targets are missed from the spectroscopy sample), we simply use the spectroscopic sample as the tracers, weighted by 1/*TSR* instead of applying the more complex ZADE approach, described above, that we developed for zCOSMOS. We also assume that the spectroscopic completeness is independent of galaxy properties in SDSS. Edge effects are treated in the same way as in zCOSMOS, but are anyway minimized by only considering objects with $f > 0.9$, where $f$ is the fraction of the adopted aperture to estimate the local density that lies within the survey region (see K10a).

For consistency with Bolzonella (2009), we define the quartiles of the environmental density using the distribution of densities of galaxies above $10^{10.5} M_\odot$.

## 3. STAR FORMATION

Star-formation represents the build-up of the visible (stellar) component of galaxies. In this section we first briefly review the strong uniformities in star-formation that have emerged from recent studies of large numbers of galaxies, both locally and at high redshifts. We then examine how these relations vary with environment, before considering the mass-function of star-forming galaxies and its evolution with epoch.

### 3.1 Star-formation rates and stellar mass

Several recent studies have emphasized the close relationship between the star-formation rates of galaxies and their existing stellar mass, $m$, conveniently parameterized as the specific star-formation rate, sSFR, defined as $sSFR = SFR/m$. In local SDSS samples, Salim et al. (2007) and Elbaz et al. (2007) have shown the existence of a tight "main sequence" of star-forming galaxies in which the sSFR is approximately constant over more than two decades of stellar mass, with a dispersion of only 0.3 dex about the mean relation. The relationship that is derived from the stellar masses and Hα-derived star-formation rates of B04 is shown in Fig 1 for blue star-forming galaxies. The ridge line of this SDSS relation has the following relation log sSFR = –10.0 – 0.1 (log $m$ – 10.0) indicating only a weak dependence of sSFR on mass, i.e. sSFR $\propto m^\beta$ with $\beta$ = -0.1. Naturally, the inverse of the sSFR defines a timescale for the formation of the stellar population of a galaxy, $\tau$ = sSFR$^{-1}$. In the local Universe, this is of order 10 Gyr, i.e. comparable to the Hubble time.

This uniformity in the sSFR in "normal" star-forming galaxies is a striking feature of the galaxy population. It clearly, however, does not extend to the Ultra Luminous Infrared Galaxies(ULIRGs) which exhibit highly elevated star formation rates of 100 $M_\odot yr^{-1}$, or greater (Sanders & Mirabel 1996) in galaxies within the same range of stellar mass of normal galaxies. However, the ULIRGs are believed to be associated with rare major mergers (Sanders et al., 1988, Sanders & Mirabel, 1996) and consequently distinct star formation processes. Although ULIRGs lie off the main sequence, their effect is in fact automatically incorporated into our analysis (as argued in Section 7.3 below) and their effect does not need to be considered separately.

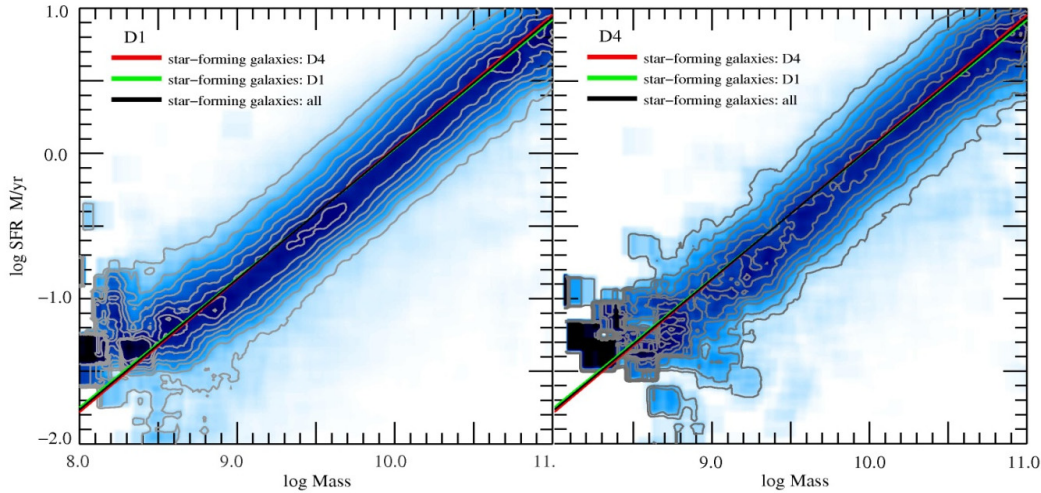

Fig 1: The relationship between SFR and stellar mass for star-forming SDSS galaxies in the low density D1 quartile (left) and high density D4 quartile (right). The three almost indistinguishable lines, reproduced on both panels, show the fitted relation for all galaxies, and for those in the D1 and D4 density quartiles. Star-forming galaxies have an sSFR that varies only very weakly with mass and is independent of environment.

The approximate constancy of sSFR with stellar mass in star-forming galaxies has also been seen at higher redshifts, e.g. Elbaz et al. (2007) at $z \sim 1$ and Daddi et al. (2007) at $z \sim 2$, and a similar relation was derived by Pannella et al. (2009) using a completely independent indicator of SFR (radio stacking). We believe that contrary results in the literature (see e.g. Maier et al., 2009) can often be ascribed to the inclusion of quenched very-low-SFR galaxies, to the use of star-formation indicators that are more sensitive to the presence of dust, or to the selection of the sample, since an SFR-selected sample will generally produce a flattening of the sSFR-$m$ relation.

In our analytic analysis below, we will follow the β dependence exactly. A constant sSFR (at a given epoch), i.e. β close to zero, is a good working hypothesis that we will adopt in our numerical simulations. Our conclusions do not actually depend on the accuracy of this assumption, and in fact our analysis provides some independent support for this hypothesis - e.g. the fact that the faint end slope of the mass-function of star-forming galaxies does not change with redshift is a natural consequence of a very weak dependence of sSFR on galactic stellar mass.

### 3.2 Independence of specific star formation rate and environment

The dependence of the star formation rate on environment has not been so well explored. The two panels of Fig. 1 show the SDSS data from B04 split into the lowest (D1) and highest (D4) density quartiles of our SDSS density field constructed as described in Section 2.2.3. There is no detectable difference between the sSFR-mass relation *for star-forming galaxies* between the two environments. This is further shown in Fig. 2, which shows the mean <log (SFR)> as a function of galactic mass and environment in the B04 sample.

This invariance of the mean sSFR on environment should not be confused with the clear evidence (see Section 4 below) that the fraction of galaxies that are star forming *does* depend quite strongly on this same environmental measure, leading to a strong environment dependence of the average SFR for the *overall* population. This distinction emphasizes that the quenching of galaxies leading to the red sequence is a relatively sharp transition. Those galaxies that are not quenched evidently continue forming stars at the same rate, regardless of their environment, despite the fact that the chance of having been quenched evidently does depend strongly on the environment.

The same invariance of sSFR with environment is seen in the zCOSMOS 10k data to $z \sim 1$. This is shown in Fig 3. Although this set of measurements is less complete and will be somewhat biased because of (mass-dependent) reddening, etc., we find no statistically significant dependence of the mean sSFR (of star-forming galaxies) on environment in zCOSMOS to $z \sim 1$.

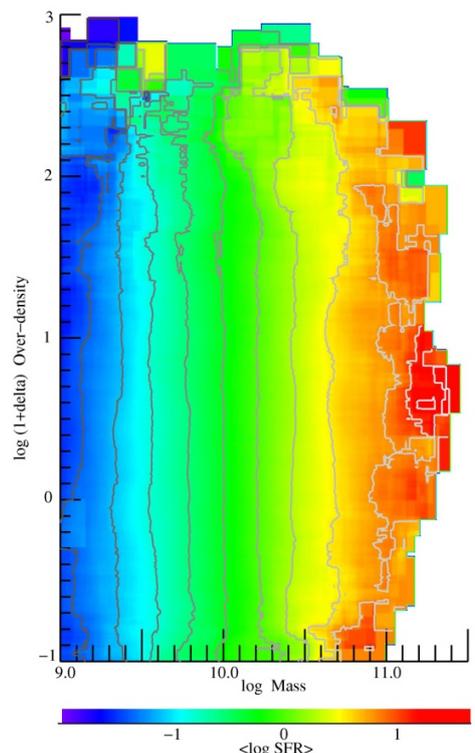

Fig 2: The mean <log SFR> as functions of stellar mass and environment for star-forming galaxies in SDSS, showing the independence of SFR on environment at given mass.

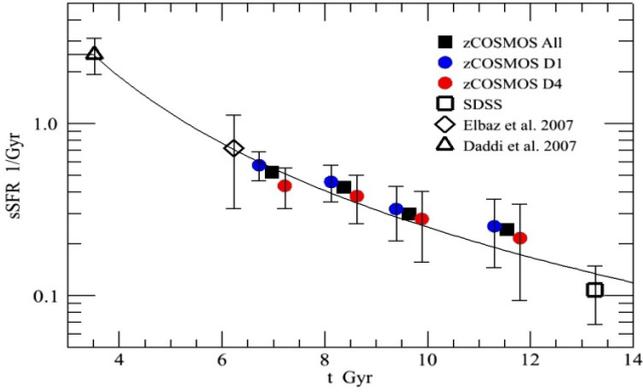

Fig 3: The mean sSFR (at masses of $10^{10}$ M$_\odot$) for blue star-forming galaxies as a function of epoch from SDSS and zCOSMOS, with values from the literature (Elbaz et al., 2007, Daddi et al., 2007). The zCOSMOS points are also split into highest and lowest quartiles of density, off-set from each other for clarity, showing an insignificant dependence of sSFR on environment in zCOSMOS.

### 3.3 Specific star-formation rate and time

It has also recently become clear that the uniformity of the sSFR described in the two previous subsections, is also seen at much higher redshifts, (e.g. Daddi et al., 2007, Elbaz et al., 2007, Dunne et al. 2009; Pannella et al. 2009; Santini et al. 2009), but with a characteristic sSFR that is substantially elevated, by a factor of about 7 at $z = 1$ and by about 20 at $z = 2$. At $z \sim 2$, the characteristic timescale $\tau$ = sSFR$^{-1}$ has thus fallen to about 0.5 Gyr, about 7 times shorter than the Hubble time at that redshift. Despite the fact that individual star formation rates have reached those associated with ULIRGs in the local universe, it is clear that these galaxies are undergoing roughly steady star-formation and are not associated with a short-lived burst of star-formation associated with a merger event. Correspondingly, the extremely luminous sub-millimeter galaxies at these redshifts have elevated star formation rates above 1000 M$_\odot$yr$^{-1}$ and once again lie off the relation defined by "normal" star-forming galaxies (Daddi et al., 2007).

Linking together the sSFR at $z \sim 2$ and $z \sim 1$ of Daddi et al. (2007) and Elbaz et al. (2007) with the zCOSMOS data (see Fig 3), and the SDSS relation above, we adopt for the <sSFR> out to $z \sim 2$ (see also Panella et al., 2009), and incorporating the mass-dependence discussed above,

$$sSFR(m,t) = \frac{1}{\tau(m,t)} = 2.5 \left(\frac{m}{10^{10} M_\odot}\right)^\beta \left(\frac{t}{3.5\,\text{Gyr}}\right)^{-2.2} \text{Gyr}^{-1} \quad (1)$$

Beyond $z \sim 2$, it appears that the characteristic sSFR flattens out and is roughly constant back to $z \sim 6$ (e.g. Gonzalez et al., 2009). The cause of this apparent change in behavior around $z \sim 2$ is not well understood but is largely incidental to our discussion.

It should be noted that this change in sSFR is responsible for the evolution in the overall star-formation rate density (SFRD) in the Universe back to these redshifts, which Lilly et al. (1996) parameterized as $t^{-2.5}$. The dramatic change in SFRD back to $z \sim 2$ is evidently not caused by an increase in the typical masses of star forming galaxies, nor by an increase in the number of star forming galaxies, since these change little with redshift (see Section 3.4 below), but rather by the large and uniform change in the SFR at a given galactic mass across the broad population of star forming galaxies.

The similarity of the exponents suggests that the overall

Table 1
The mass-function of star-forming galaxies (adapted from Ilbert et al., 2010)

| $z$ | M* | $\alpha_s$ | $\phi$*/$10^{-3}$Mpc$^{-3}$ |
|---|---|---|---|
| (a) Free-fits | | | |
| 0.3 | 10.99 | -1.31 | 1.21 |
| 0.5 | 11.02 | -1.30 | 0.75 |
| 0.7 | 10.96 | -1.35 | 0.80 |
| 0.9 | 10.89 | -1.22 | 1.22 |
| 1.1 | 10.94 | -1.24 | 0.84 |
| 1.35 | 10.89 | -1.26 | 0.74 |
| 1.75 | 10.94 | -1.26 | 0.48 |
| (b) Fits with $\alpha_s$ constrained | | | |
| 0.3 | 10.97 | (-1.30) | 1.28 |
| 0.5 | 11.02 | (-1.30) | 0.75 |
| 0.7 | 10.90 | (-1.30) | 0.98 |
| 0.9 | 10.96 | (-1.30) | 0.89 |
| 1.1 | 11.00 | (-1.30) | 0.67 |
| 1.35 | 10.95 | (-1.30) | 0.62 |
| 1.75 | 10.99 | (-1.30) | 0.41 |

Notes to Table: These stellar mass values should not be compared directly with those used elsewhere in the paper.

evolution of the sSFR at $z < 2$ reflects the evolution of the specific accretion rate of haloes (in both baryonic and dark matter) as they hierarchically grow. The flattening at high redshifts is poorly understood but may reflect a limit to the sSFR.

### 3.4 The mass-function of star-forming galaxies

The mass (and luminosity) function(s) of blue star-forming galaxies can be well fit by a Schechter function (e.g. Lilly et al., 1995, Bell et al., 2003, 2007, Ilbert et al., 2005, Zucca et al., 2006). Surprisingly, it has become increasingly clear that the shape of the mass-function stays remarkably constant over a large range of redshifts, despite the large increase in the masses of individual galaxies implied by the star-formation law given in equation (1). The characteristic mass M* and faint end slope $\alpha_s$ stay essentially constant, whereas the overall normalization $\phi$* drifts upwards with time especially at high redshifts $z > 1$. This constancy is clearly seen in the spectroscopic zCOSMOS sample to $z \sim 1$ (see Fig 12 of Pozzetti et al., 2009) and in the photo-z COSMOS sample to $z \sim 2$ (see Fig 18 of Ilbert et al., 2010), and has previously been remarked upon by others, including Bell et al. (2007, see their Fig 1).

An example of this remarkable fact is shown in Table 1 which shows the Schechter parameters obtained by re-fitting a single Schechter function to the sum of the components that are identified in the Ilbert et al. (2010) COSMOS analysis with "intermediate-" and "high-" activity galaxies, i.e. omitting the "quiescent population". Over the whole range $0.1 < z < 2$, both M* and $\alpha_s$ remain essentially constant within 0.05 dex and 0.05 respectively, within this highly homogeneous data set (thereby avoiding any issues of mass determination from sample to sample). The normalization $\phi$* is more variable, partly due to large scale structure in COSMOS, but is more or less constant to $z \sim 1$, but then declines by a factor of about three to $z \sim 2$. Pérez-González et al. (2008) also constructed a mass-function to $z \sim 4$ which shows the same behavior, i.e. constant M* and $\alpha_s$ and with $\phi$* slowly increasing with time, especially at $z > 1$.

Individual star-forming galaxies will be increasing their masses through the star-formation described by equation (1). Integration of the sSFR relation over time indicates that a galaxy which is not quenched and which remains on the blue

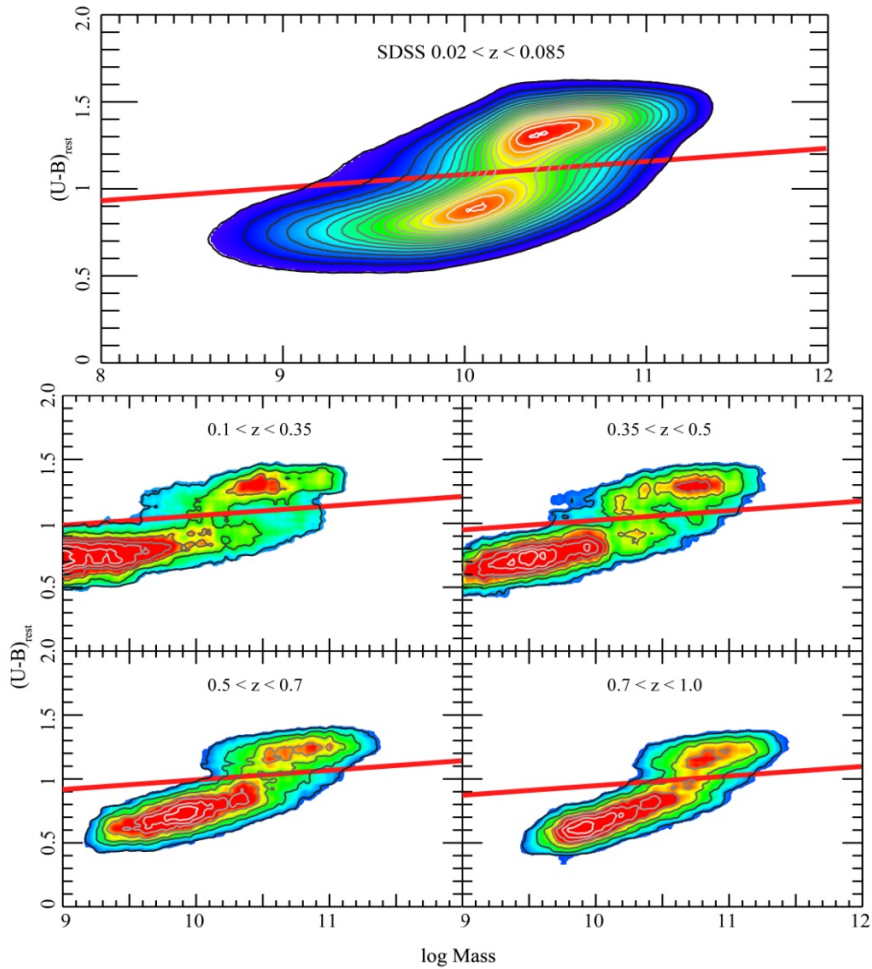

Fig 4: Color distributions in SDSS (upper) and zCOSMOS at different redshifts (lower panels) with the dividing line used to split galaxies into red and blue.

"main sequence" will have increased its mass by about 2.0 dex since $z = 2$ (i.e. $\Delta \log m = +2.0$) and by 0.75 dex since $z = 1$, despite the rapid fall in sSFR. These star-forming galaxies are clearly a mobile population of galaxies moving steadily through the mass function, emerging to be quenched at the high mass end.

This paper is primarily concerned with the "quenching" of galaxies as a function of cosmic epoch, galactic stellar mass, and environment. Quenching is therefore distinct from the smooth and uniform cosmological evolution in the sSFR that is given by equation (1). To a large degree the precise time evolution of the sSFR is incidental to our analysis and serves to set the "cosmic clock" for the process (see Section 7.2 below), rather than controlling the eventual outcome.

## 4. QUENCHING

We consider the term "quenching" to mean the cessation of star-formation, for any reason. Operationally, in our analysis, quenching is the process that causes the color change associated with the migration of galaxies from the so-called "blue cloud", where the star-formation continues as described in Section 3, to the "red sequence", where it is assumed star-formation activity is so suppressed as to be negligible. We recognize that this may be an over-simplification since some highly reddened star forming galaxies will have red colors and will masquerade on the red sequence. Quenching may be either internally triggered, or externally triggered. We will also assume for simplicity that the merging of two galaxies also leads to a cessation of star formation.

Star-formation may cease in an isolated galaxy, which thereafter remains at the same mass. We denote this quenching rate as $\lambda$. Star-formation is also assumed to cease after a major merger event, which will discontinuously change both the number and masses of the galaxies concerned. We denote the major merger rate as $\kappa$. For some purposes the distinction is unimportant, and we will not surprisingly encounter observational degeneracies between these two processes. Both result in the "death" of the galaxy and we will sometimes consider the combined death rate, $\eta$, which will be the sum of $\lambda$ and $\kappa$. These three parameters $\eta, \kappa, \lambda$, all have the dimensions of $time^{-1}$ and reflect the probability that a particular galaxy will be quenched in unit time. Subsequent "post-quenching" merging may also modify the masses and number densities within the population of already-quenched galaxies, and this is considered separately in Section 5.4.

We also note that some passive galaxies may be "revived" by restarting star-formation. However, this detail is largely immaterial, provided that the quenching rate is interpreted as the "net" quenching from blue to red.

Finally, we comment that we will for simplicity consider quenching to be an instantaneous event. When considering quenching "timescales", we will be referring to the time that a galaxy statistically waits to be quenched (i.e. the inverse of the quenching rate), rather than the timescale for the actual physical quenching transformation to take place, from start to completion, which we assume is very short.

In the local Universe the fraction of galaxies that are on the red sequence, $f_{red}$, is a function of both mass (e.g. Kauffmann et al., 2003a) and environment (e.g. Kauffmann et al., 2004, Baldry et al., 2006). For simplicity, we consider only two states for galaxies, "blue star forming" and "red passive", based on a dividing rest-frame (U-B) color, which is a weak function of mass and which will drift to bluer colors at higher redshifts. This is obviously somewhat simplistic, but is in the spirit of our approach to identify the most basic features of the galaxy population. Fig 4 shows this division both for SDSS and zCOSMOS. The dividing lines that we adopt are given as follows:

$$(U-B)_{AB} = 1.10 + 0.075 \log\left(\frac{m}{10^{10} M_\odot}\right) - 0.18z \quad (2)$$

*4.1 Formalism: differential effects of mass and environment*

At any epoch and in any environment, the fraction of galaxies in the red and blue "states" are given by $f_{red}$ and $f_{blue}$, which sum to unity. Empirically, the fraction $f_{red}$ in the local Universe is found to increase with galaxy stellar mass and with environmental density (see Fig 11 of Baldry et al., 2006).

To compare environments at different epochs we consider the over-density δ discussed in Section 2.1.3. This is equivalent to the comoving density ρ and we will use these interchangeably. It might be thought that the physical density would be a more useful quantity. However, our environmental scale of order 1 Mpc is outside of the virial radius of most haloes. The comoving density on this comoving scale best reflects broad environmental differences between voids, filaments and clusters, which appear to control some properties of dark matter haloes (see e.g. Section 4.3 below).

At a given epoch, we define the *relative* environmental quenching efficiency, $\varepsilon_\rho$, as follows: $\varepsilon_\rho$ is the fraction of those galaxies at a given galactic mass, $m$, which would be forming stars (i.e. be blue) in some reference environment, $\rho_0$, but which are however progressively quenched (i.e. are red) in denser environments:

$$\varepsilon_\rho(\rho, \rho_0, m) = \frac{f_{red}(\rho, m) - f_{red}(\rho_0, m)}{f_{blue}(\rho_0, m)} \quad (3)$$

It is convenient to choose $\rho_0$ to be the lowest density environment, i.e. the most void-like regions, where one might expect environmental effects to be minimum and where the dependence of galactic properties with environment is in any case seen to saturate (Baldry et al., 2006). This choice will henceforth be implied, so that $\varepsilon_\rho$ will be always positive, and never larger than unity.

It is important to stress that $\varepsilon_\rho$ measures the *differential* quenching effect of the environment *starting* from the population of galaxies (at the same stellar mass) that is seen in the lowest density regions. The point of normalizing the change in the color distribution of the galaxy population in this way is that it makes $\varepsilon_\rho$ insensitive to the addition of extra red galaxies provided that the size of this additional component is independent of environment.

There is also an equivalent relative mass quenching efficiency, $\varepsilon_m$, which measures the differential effect of stellar mass in determining the red fraction at some fixed environmental density.

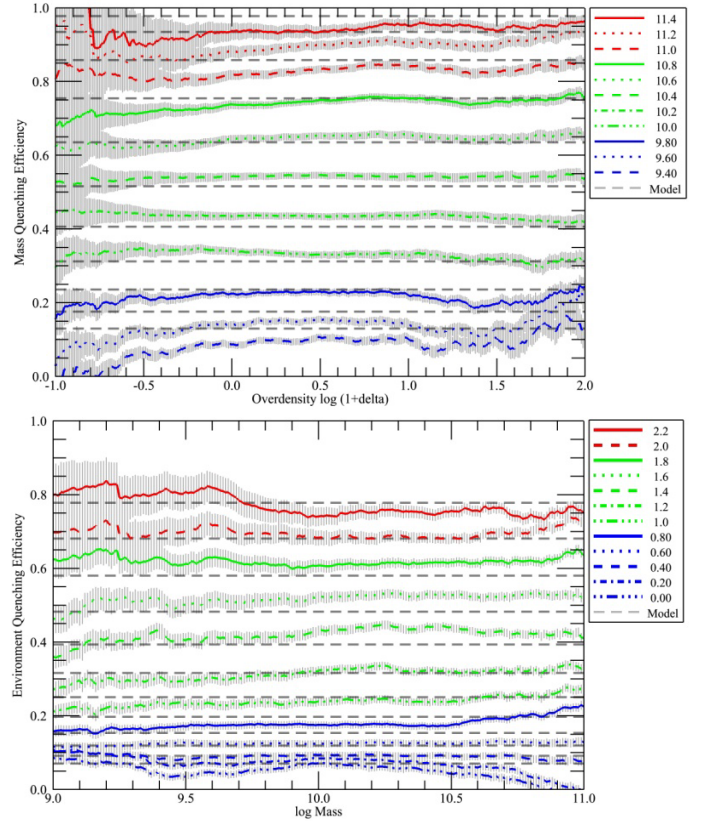

Fig 5: Observed values of the relative mass quenching efficiency, $\varepsilon_m$, as a function of environment for different galaxy masses (top) in units of log solar mass, and of the relative environment quenching efficiency, $\varepsilon_\rho$, as a function of mass for different environments (bottom) in units of log (1+δ). The fact that these are essentially flat shows that the differential effects of mass and environment are separable in SDSS. The model curves are for the simple parameterizations given in the text and Table 2.

$$\varepsilon_m(m, m_0, \rho) = \frac{f_{red}(m, \rho) - f_{red}(m_0, \rho)}{f_{blue}(m_0, \rho)} \quad (4)$$

It is again convenient to set the reference mass $m_0$ to a very low mass where almost all galaxies (at least in the voids) are blue.

In this most general formalism both $\varepsilon_m$ and $\varepsilon_\rho$ may be functions of both $m$ and $\rho$. However, we show in the next section that in fact $\varepsilon_m$ is independent of $\rho$, and $\varepsilon_\rho$ is independent of $m$.

*4.2 The empirical separability of environment and mass*

Fig 5 shows the empirical values of $\varepsilon_m$ and $\varepsilon_\rho$ as functions of mass and environment in the SDSS sample. These are determined within moving boxes of size 0.3 dex in mass and 0.3 dex in (1+δ). The grey hatched regions around selected lines show typical observational (sampling) uncertainties which have been simply derived from the binomial error of the fraction in the box (68% confidence level).

To a remarkable degree, the relative mass quenching efficiency function, $\varepsilon_m$, is found to be *independent* of environment, and the relative environmental quenching efficiency function $\varepsilon_\rho$ is found to be *independent* of galactic stellar mass. The horizontal lines in the two panels of Fig 5 show the fitted relations.

$$\varepsilon_\rho(\rho, \rho_0) = (1 - \exp(-(\rho/p_1)^{p_2}))$$
$$\varepsilon_m(m, m_0) = (1 - \exp(-(m/p_3)^{p_4})) \quad (5)$$

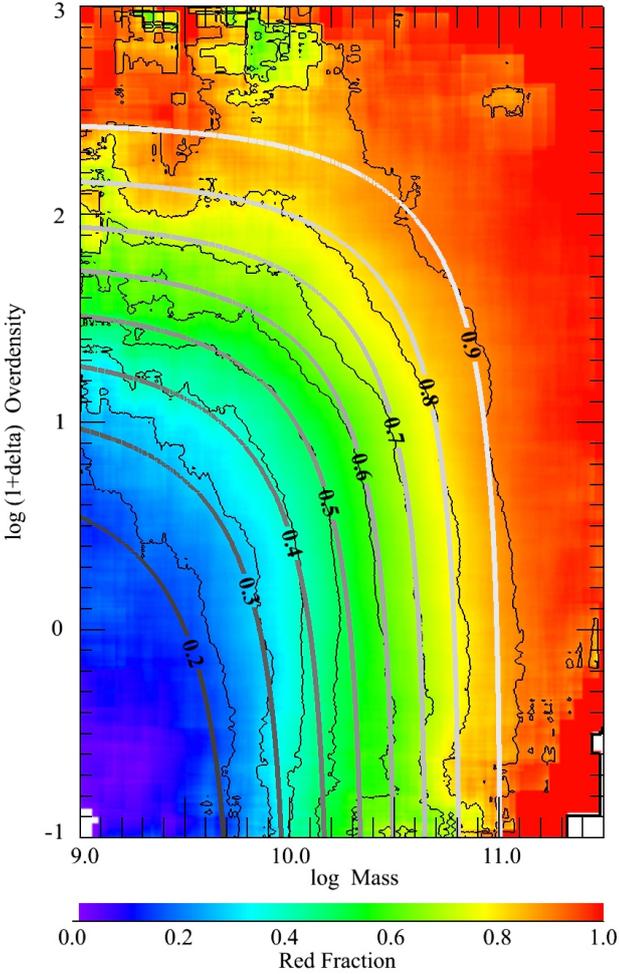

Fig 6: The red fraction in SDSS as functions of stellar mass and environment.

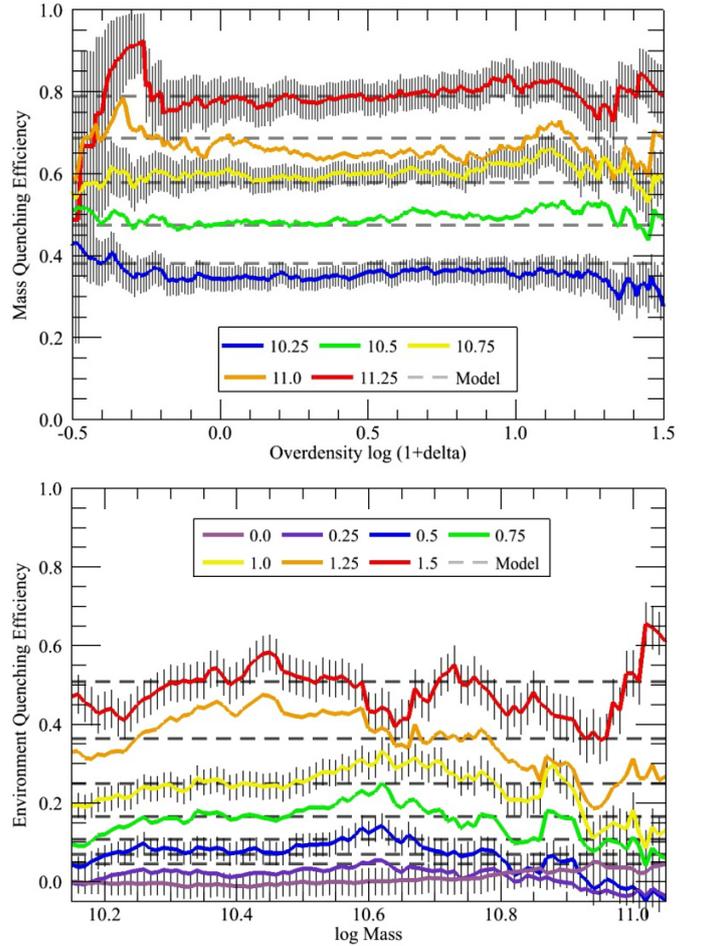

Fig 7: As for Figure 5, but for the zCOSMOS sample at $0.3 < z < 0.6$.

with the values $p_1$ to $p_4$ given in Table 2, plotted at intervals of 0.2 dex in $m$ and $\rho$.

The separation of the effects of mass and environment is naturally not perfect but holds over two orders of magnitude in both mass and environmental density, with local deviations from the horizontal lines that are comparable to the observational uncertainties. The limited excursions of the data show that deviations from this simple separable behavior in $m$ and $\rho$ are rather small, equivalent to no more than ±0.2 dex in either variable, a tenth or less of the overall range of each parameter.

In other words, the differential effect of the environment on the red-blue mix of galaxies in SDSS is independent of galactic stellar mass, and vice versa. This good empirical separability of mass and environment means that we can write the red fraction in terms of $\varepsilon_m$ and $\varepsilon_\rho$, by either of the first two equations, which reduce to the third:

$$\begin{aligned} f_{red}(\rho,m) &= \varepsilon_m(m,m_0) + \varepsilon_\rho(\rho,\rho_0)[1-\varepsilon_m(m,m_0)] \\ &= \varepsilon_\rho(\rho,\rho_0) + \varepsilon_m(m,m_0)[1-\varepsilon_\rho(\rho,\rho_0)] \\ &= \varepsilon_\rho + \varepsilon_m - \varepsilon_\rho\varepsilon_m \end{aligned} \quad (6)$$

with $\varepsilon_m$ independent of $\rho$ and with $\varepsilon_\rho$ independent of $m$. This implies a simple symmetry to the $f_{red}(\rho, m)$ surface, which is illustrated in Fig 6.

Since $\varepsilon_\rho$ is zero in the lowest density regions (i.e. the voids), this separability means that $\varepsilon_m(m)$ is easily interpreted as the red fraction in these lowest density regions. Likewise, $\varepsilon_\rho(\rho)$ is the red fraction for very low mass galaxies, for which $\varepsilon_m$ is by construction zero.

By inserting the two fitted relations (5) into (6), we recover

$$f_{red}(\rho,m) = 1 - \exp\left(-(\rho/p_1)^{p_2} - (m/p_3)^{p_4}\right) \quad (7)$$

which was previously proposed by Baldry et al. (2006) as one of two empirical fitting functions for the $f_{red}(\rho, m)$ surface in SDSS.

The clear separability of the effects of environment and mass, when parameterized in this way, suggests that there are two distinct processes at work. We will henceforth refer to these as "environment-quenching" and "mass-quenching" to reflect their (independent) effects on $f_{red}$ across the $(\rho, m)$ plane. These two quenching processes will be governed by rates (i.e. the probability of being quenched per galaxy per unit time) of $\lambda_\rho$ and $\lambda_m$ respectively.

The distinction between the two effects will be even more clearly seen when we consider how, observationally, $\varepsilon_m$ and $\varepsilon_\rho$ depend on cosmic epoch. For this we turn to our zCOSMOS sample in the next Section.

*4.3 How the environment-quenching operates*

*4.3.1 The empirical signature of environment-quenching*

Fig 7 shows the equivalent plots of $\varepsilon_m$ and $\varepsilon_\rho$ from the

Table 2: Best fit parameters for relative quenching efficiencies

| sample | log ($p_1$) [a] | $p_2$ | log ($p_3$) [b] | $p_4$ |
|---|---|---|---|---|
| SDSS DR7 0.02< z <0.085 | 1.84 ±0.01 | 0.60 ±0.01 | 10.56 ±0.01 | 0.80 ± 0.01 |
| zCOSMOS 0.10 < z < 0.35 | ... | ... | 10.78 ± 0.02 | 0.78 ± 0.02 |
| zCOSMOS 0.35 < z < 0.50 | 1.86 ± 0.03 | 0.67 ± 0.09 | 10.76 ± 0.03 | 0.61 ± 0.03 |
| zCOSMOS 0.50 < z < 0.70 | 1.74 ± 0.04 | 0.70 ± 0.11 | 10.83 ± 0.03 | 0.65 ± 0.04 |
| zCOSMOS 0.70 < z < 1.00 | 1.90 ± 0.04 | 0.64 ± 0.15 | 10.89 ± 0.05 | 0.63 ± 0.06 |

Notes to Table
[a] in units of $1+\delta_5$
[b] in units of $M_\odot$

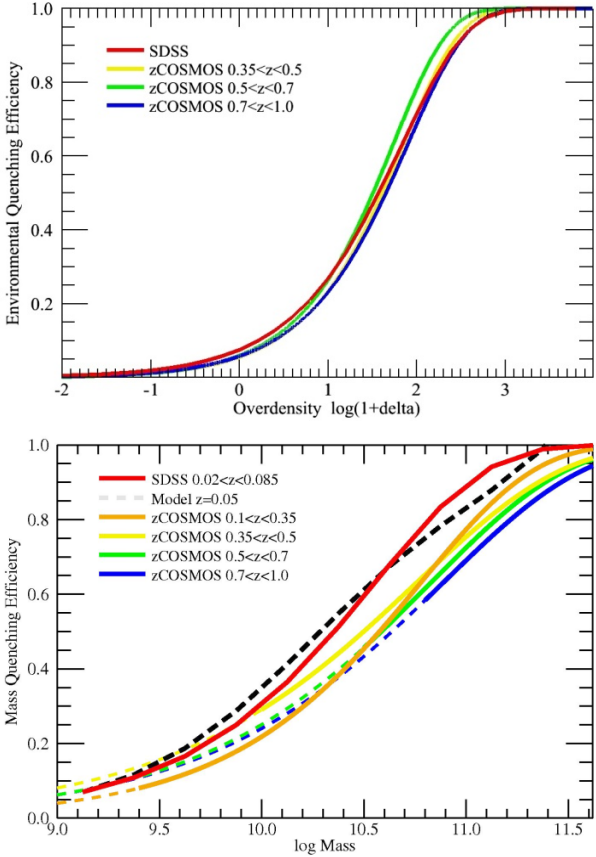

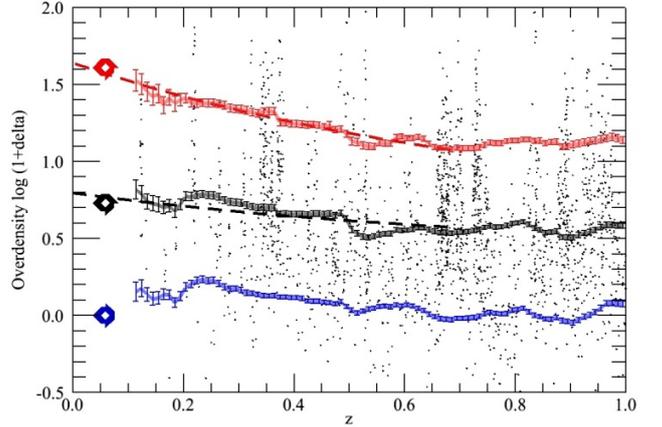

Fig 9: The growth of structure in zCOSMOS and SDSS. The median overdensity of all galaxies (black) and of those in the lowest (blue) and highest (red) density quartiles is plotted as a function of redshift. As the denser environments get progressively denser, galaxies migrate to the higher values of the relative environmental quenching efficiency $\mathcal{E}_\rho$ shown in the upper panel of Fig 8. The red and black dashed curves show the fitted relations to the zCOSMOS and SDSS data. In D1, where $\delta \sim 0$, it is assumed that the growth of structure is negligible.

Fig 8: Comparison of the relative environmental quenching efficiency $\mathcal{E}_\rho$ (top) and the relative mass quenching efficiency $\mathcal{E}_m$ (bottom) in SDSS and in three zCOSMOS redshift bins. Dashed extensions show mass regimes that are not directly constrained by the data. The differential effect of environment is essentially independent of epoch in the upper panel, whereas the differential effect of mass changes with time in the lower panel. The dashed black line in the lower panel is the $z = 0$ prediction of our very simple quenching law based on star-formation, which it should be noted is derived completely independently of the plotted (observational) lines.

zCOSMOS-10k data for 0.3 < z < 0.6. Although the zCOSMOS data is inevitably noisier, a good degree of separation between the effects of mass and environment is again discernable in zCOSMOS. The horizontal lines again show the fits to functions of the form of (5), with the values given in Table 2.

Fig 8 then compares the form of $\mathcal{E}_m(m)$ and $\mathcal{E}_\rho(\rho)$, derived from zCOSMOS in three redshift bins, using a fitting function of the form of equation (5), with the values given in Table 2. We omit the lowest redshift bin (0.1 < z < 0.35) because it suffers from both a small overall volume and more severe edge corrections to the densities, and is dominated by a few large structures (see Fig 9 below). It is clear that the environmental efficiency curve $\mathcal{E}_\rho$ does not vary significantly with redshift, neither within the zCOSMOS data set, nor between zCOSMOS and the local SDSS sample. Evidently, the differential effects of the environment, at fixed over-density, are independent of both stellar mass and cosmic time.

It is important to appreciate that the redshift-independence of our environmental quenching efficiency parameter $\mathcal{E}_\rho(\rho)$ does *not* imply that the effect of the environment on the galaxy population is unchanging, since the environments of almost all galaxies will be migrating to higher over-densities as large scale structure develops through gravitational instability, since almost all galaxies occupy regions with $\delta > 0$. This growth of structure is seen in Fig 9, where we split the zCOSMOS galaxy population into quartiles of density and plot the median density of the highest and lowest quartiles as a function of redshift. The median over-density of a given quartile increases steadily with cosmic epoch, and does so at a faster rate for the higher density environments. The galaxy population therefore occupies a shifting locus in $\rho$ on the unchanging $\mathcal{E}_\rho(\rho)$ curve, progressively broadening in $\rho$ and extending further up onto the steeper part of the $\mathcal{E}_\rho(\rho)$ curve as time passes.

Environmental effects within the galaxy population therefore develop and accelerate over time as the galaxy population migrates to a broader range of densities. This can be seen in our earlier zCOSMOS analyses of $f_{red}$ in Cucciati et al. (2009), and the analogous analysis of morphology in Tasca et al. (2009), in which we split the galaxy population by environmental density quartiles as here. Both analyses showed a progressive development of differences between the highest and lowest density quartiles as the redshift decreased from $z \sim 1$ to locally. Environmental effects are much weaker at $z \sim 1$ than today simply because many fewer galaxies inhabit the high $\delta$ regions where the (unchanging) environmental effects are strongest.

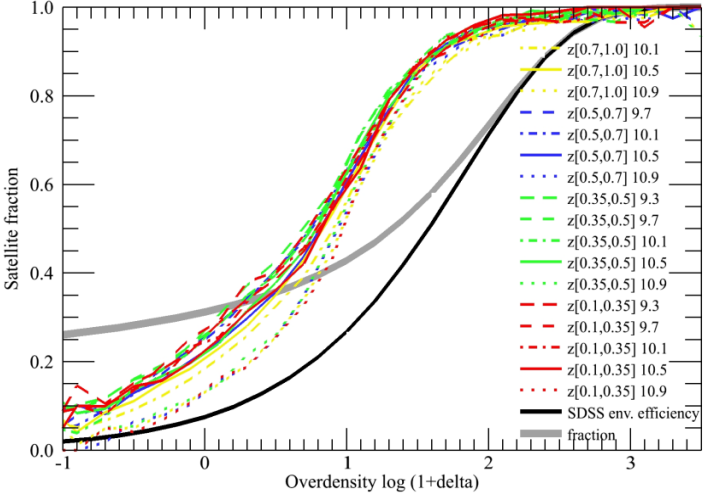

Fig 10: The satellite fraction for different stellar masses as a function of environmental over-density in the COSMOS mocks of Kitzbichler and White (2007), based on the Millenium simulation, in the four redshift bins used for our zCOSMOS analysis. The curves show very little dependence on galactic mass, especially below $m \sim 10^{10.9}$ M$_\odot$, and virtually no dependence on redshift. These are essentially the same properties as for our relative quenching efficiency, $\mathcal{E}_\rho$, making satellite quenching an obvious and attractive candidate for our environmental-quenching process. The black curve shows our $\mathcal{E}_\rho$ function, which is the fraction of blue galaxies that have been quenched by the environment for very low mass galaxies, and the grey curve shows the implied fraction of satellites that have been quenched. This varies between 30-70% of the satellites for over-densities with log $(1+\delta) < 2$, where most of the galaxies reside.

### 4.3.2 Physical origin of environment-quenching

In the previous sub-section, we showed that the environment apparently imprints itself on the galaxy population in a way that is uniquely given by the environment (over-density), *independent* of epoch *and* of the mass of the galaxy.

A natural contender for this characteristic of the environmental effect is the quenching of galaxies as they fall into larger dark matter haloes (Larson et al.1980; Balogh, Navarro & Morris 2000; Balogh & Morris 2000; van den Bosch et al. 2008). Examination of the 24 COSMOS mock catalogues (Kitzbichler & White 2007) shows that the fraction of galaxies, at a given mass, that are satellite galaxies, $f_{sat}$, is strongly environment dependent, but, at fixed $\rho$ or $\delta$, is almost entirely independent of epoch, (at least since $z = 0.7$) and of galactic mass, $m$ (especially for $m < 10^{10.9}$ M$_\odot$), as shown in Fig 10. These are precisely the same two characteristics that we have identified empirically for our "environment-quenching" process.

If we suppose that "satellite-quenching" quenches some fraction $x$ of satellites as they fall into larger haloes, then it is easy to see that $x$ will be given by the ratio of $\mathcal{E}_\rho/f_{sat}$. Inspection of Fig 10 shows that $x$ takes a value that increases from about 30% at the lowest densities up to about 75% for our densest environments with log $(1+\delta)$ ~2. Interestingly, this is in the same range as the estimate (40%) of the fraction of satellites that are quenched from van den Bosch et al. (2008).

Ram pressure stripping and strangulation are usually considered as the physical mechanism through which satellite quenching operates (see e.g. Feldmann et al., 2010). Such processes may efficiently quench star-formation, but would not probably lead to morphological transformations. Incorporation of morphological information into our picture could help better understanding this process, but this is beyond the scope of the current paper.

### 4.4 How mass-quenching operates

#### 4.4.1 Numbers of galaxies

In Fig 8, the relative mass quenching efficiency $\mathcal{E}_m(m)$ changes somewhat with cosmic epoch. More importantly, the individual masses of star-forming galaxies will be continuously growing due to their star-formation. From both points of view, mass-quenching must be a continuous process, and we therefore now consider the transformation rates of galaxies in this mass-quenching process.

It is convenient to approach this via consideration of the numbers of galaxies. For a given set of galaxies, we define the transformation rate, $\lambda$ as the fraction of isolated (non-merging) blue galaxies that are quenched to form red galaxies per unit time. This may well be a function of mass and/or environment, and will have components due to both the mass-quenching and environment-quenching, $\lambda_m$ and $\lambda_\rho$. We have seen above that the $\lambda_\rho$ term comes from the change in $\rho$ with redshift, rather than the change in $\mathcal{E}_\rho$.

We must also consider the merging of galaxies, which we assume involves both blue and red galaxies (proportionally) as progenitors but which produces only red galaxies. For several reasons, it is convenient to characterize the merging rates $\kappa$ in terms of the number of galaxies at the galactic mass of interest (rather than at the mass of the progenitors). We will therefore write the influx of new red galaxies *into* the mass bin per unit time as the product $\kappa_+ N_{tot}$, while the efflux of red and blue galaxies *out* of the mass bin per unit time is assumed to be $\kappa_- N_{red}$ and $\kappa_- N_{blue}$ respectively. The merging rates $\kappa$ may well be functions of environment (Lin et al., 2010, de Ravel et al., in preparation, Kampczyk et al., in preparation) and possibly also of galactic stellar mass.

The number of blue galaxies in a given (fixed) mass bin will also be changing simply because ongoing star-formation will bring up lower mass galaxies into the mass bin in question. Since the sSFR is roughly constant, the shift in logarithmic mass will be at most only weakly dependent on mass. We can therefore characterize this increase in the numbers via the logarithmic slopes of the mass-function of star-forming galaxies and the sSFR-mass relations, the latter acting to compress more galaxies into a given interval of log mass. We henceforth define

$$\alpha = \frac{d\log\phi_{blue}}{d\log m}$$
$$\beta = \frac{d\langle\log sSFR\rangle}{d\log m} \quad (8)$$

with $\phi_{blue}$ the number density of blue galaxies per unit log interval in mass. With this definition of b, the SFR is proportional to $m^{1+\beta}$. It should be noted that, observationally, $\beta$ is close to zero, but is probably slightly negative, $\beta \sim -0.1$. We will follow $\beta$ through in the following analysis, but will often assume for simplicity that it is zero. In the end our conclusions are not very dependent on the precise value of $\beta$.

The parameter $\alpha$ will also be negative, strongly so at high masses, since for the well-known Schechter (1976) function given by $\phi(m) = \phi^*(m/M^*)^{1+\alpha_s}\exp(-m/M^*)$, $\alpha$ will be given by

$$\alpha = (1+\alpha_s) - \frac{m}{M^*} \quad (9)$$

Considering the changes in the numbers of red and blue galaxies, we get the following equations for the rate of change in the number of galaxies $N$ per unit logarithmic mass bin, at fixed mass and environment,

$$\left.\frac{dN_{red}}{dt}\right|_{m,\rho} = N_{blue}\lambda_m + \kappa_+ N_{tot} - \kappa_- N_{red}$$
$$\left.\frac{dN_{blue}}{dt}\right|_{m,\rho} = N_{blue}\{-(\alpha+\beta)sSFR - \lambda_m - \kappa_-\} \quad (10)$$
$$\left.\frac{dN_{tot}}{dt}\right|_{m,\rho} = -N_{blue}(\alpha+\beta)sSFR + (\kappa_+ - \kappa_-)N_{tot}$$

These exact continuity equations simply reflect the definitions of the quenching and merging rates $\lambda$ and $\kappa$, once separability of the effects of mass and environment is adopted. The $\alpha \times sSFR$ term accounts for the change in numbers of galaxies due to their increase in mass (bringing up galaxies from further down the star-forming mass-function), while the $\beta \times sSFR$ term accounts for the compression or expansion of a given interval of logarithmic mass if $\beta \neq 0$. They are valid at fixed $m$ and $\rho$, which is why $\lambda_\rho$ does not appear (since we have argued that environment-quenching occurs as galaxies change environment). Later we will use the equation for $dN_{blue}/dt$ for the population averaged over all $\rho$, and in this case, there will be an additional $\langle\lambda_\rho N_{blue}\rangle$ term on the right-hand side of the first two equations (entering with plus and minus signs respectively).

*4.4.2 Information on mass-quenching from the red fraction*

For completeness, we look first at the change in the red fraction that is implied from these relations. However, we find that this is not as useful as a second approach (Section 4.4.3) and this section may be skipped if desired.

The change in $f_{red}$ at fixed $m$ and $\rho$, is given by

$$\left.\frac{df_{red}}{dt}\right|_{m,\rho} = \frac{1}{N_{tot}}\left.\frac{dN_{red}}{dt}\right|_{m,\rho} - \frac{N_{red}}{N_{tot}^2}\left.\frac{dN_{tot}}{dt}\right|_{m,\rho}$$
$$= (1-f_{red})\{\lambda_m + \kappa_+ + f_{red}(\alpha+\beta)sSFR\} \quad (11)$$

The effect of merging *out* of the bin $\kappa_-$ has vanished in (11), since merging was assumed to involve both blue and red galaxies proportionally. In terms of the simple red fraction, it can be seen that the effect of merging *into* the bin, $\kappa_+$, is indistinguishable from the blue to red quenching rate for isolated galaxies $\lambda_m$, and so we can lump these two together:

$$(\lambda_m + \kappa_+) = \frac{1}{(1-f_{red})}\left.\frac{df_{red}}{dt}\right|_{m,\rho} - f_{red}(\alpha+\beta)sSFR \quad (12)$$

The last term accounts for the fact that the number of blue galaxies (and therefore the red fraction) will change due to star-formation increasing the mass of the blue galaxies. This will generally bring up more low mass galaxies than are moved out of the bin, unless the $(\alpha+\beta)$ combination is zero (for example if the mass-function of blue galaxies is flat and the sSFR is independent of mass), or unless there is no star-formation.

The $df_{red}/dt$ at fixed $m$ and $\rho$ can be expressed in terms of the partial time derivatives of observed quenching efficiencies introduced earlier:

$$\left.\frac{df_{red}}{dt}\right|_{m,\rho} = (1-\varepsilon_\rho)\frac{\partial\varepsilon_m}{\partial t} + (1-\varepsilon_m)\frac{\partial\varepsilon_\rho}{\partial t} \approx (1-\varepsilon_\rho)\frac{\partial\varepsilon_m}{\partial t} \quad (13)$$

the final approximation holding since $\varepsilon_\rho$ is observed to be essentially time independent (see Section 4.3).

The environment-quenching rate $\lambda_\rho$ that occurs as objects migrate towards higher densities (and thus higher $\varepsilon_\rho$) is straightforward since the numbers of objects will be conserved.

$$(1-f_{red})\lambda_\rho = \frac{\partial f_{red}}{\partial\log\rho}\frac{\partial\log\rho}{\partial t}$$
$$= (1-\varepsilon_m)\frac{\partial\varepsilon_\rho}{\partial\log\rho}\frac{\partial\log\rho}{\partial t} \quad (14)$$

Putting together equations (13) and (6), and then (14) and (6), we therefore obtain the two transformation rates in terms of the observable quantities $\varepsilon_m$, $\varepsilon_\rho$, $f_{red}$, $\rho$, $\alpha$, and $\beta$ as follows:

$$(\lambda_m + \kappa_+) = \frac{1}{(1-\varepsilon_m)}\frac{\partial\varepsilon_m}{\partial t} - f_{red}(\alpha+\beta)sSFR$$
$$\lambda_\rho = \frac{1}{(1-\varepsilon_\rho)}\frac{\partial\varepsilon_\rho}{\partial\log\rho}\frac{\partial\log\rho}{\partial t} \quad (15)$$

The $(\lambda_m + \kappa_+)$ combination could be functions of $m$ and $t$, and also conceivably of $\rho$, since the merging rate may depend on $\rho$ and because the $f_{red}$ in the mass-growth term will also be higher in high density environments. The environmental transformation rate $\lambda_\rho$ should not have any dependence on mass because of the demonstrated separability of $\varepsilon_\rho$ and $\varepsilon_m$ noted above. The simple form of these equations is a consequence of the separability of our two evolutionary processes in the $(m, \rho)$ plane.

Given the small changes in $\varepsilon_m$ with cosmic epoch (see Fig 8), the dominant term in the quenching of galaxies $(\lambda_m + \kappa_+)$ in equation (15) is the second term, due to the increase in mass of the star-forming galaxies, and not the first, which arises from the change in red fraction at fixed mass. The precise time dependence of the $\varepsilon_m$ curve is also quite sensitive to the choice for the redshift dependence of the color cut, and it is noticeable that the curves for zCOSMOS show even less change within their redshift range than the difference between zCOSMOS and SDSS. The apparent changes in $\varepsilon_m$ in Fig 8 should therefore be treated with caution.

For both these reasons, it is therefore more revealing to look directly at the evolution of the mass-function of star-forming galaxies, which looks at the galaxies that have *not* been quenched. This topic is therefore examined in the next section.

*4.4.3 Information on mass-quenching from the mass-function of star-forming galaxies*

The mass-function of star-forming galaxies reflects the survival of those star-forming galaxies that have not been quenched, giving a clearer view of the action of quenching. Consideration of the star-forming mass-function, rather than the color mix, will require us to average the effects of environment quenching since we will lose direct information on the environment. This is almost unavoidable because the definition of environment (e.g. quartiles of the density distribution of galaxies) will be intrinsically linked to the amplitude of the observed mass- (or luminosity-) functions.

We commented above (Section 3.4) on the remarkable stability of the mass-function of star-forming galaxies over a broad range of epochs, despite the large increase in the masses of star-forming galaxies obeying equation (1) (see e.g. Renzini et al., 2009). We show in this section that the observed constancy of the exponential cut-off M* for the *star-forming* mass function demands a strikingly simple form of the mass quenching law in which the mass-quenching rate is directly proportional to the SFR alone, independent of epoch. This must operate for masses around and above M*. In Section 5.2, we show that this mass-quenching law also naturally explains the Schechter form of the passive mass-function over a much wider range of masses, establishing its viability over about two decades of galactic mass.

It is easy to see how this requirement comes about at high masses in the regime of the exponential cut-off: the rate of increase in log $m$ of a galaxy is proportional to its individual sSFR. If the sSFR across the population is at least approximately independent of mass, then the mass-function of star-forming galaxies will shift to higher masses at a speed (in log $m$ space) that is proportional to the sSFR. This shift must be counteracted by quenching, and so the quenching rate must be proportional to the sSFR. Secondly, in order to maintain the exponential cut-off at the same mass, the quenching rate must also be proportional to mass, because more rapid quenching is required where the mass-function of star-forming galaxies is steepest, i.e. where the effect of increasing mass will most strongly effect the numbers of objects. The logarithmic slope $\alpha$ of the Schechter function is proportional to mass, at high masses (see equation 9). The product $m \times $sSFR is clearly just the SFR.

We can also see this analytically by looking at what would be required *if* the mass-function of blue star-forming galaxies had a Schechter form that was observed to be *absolutely* constant with epoch, i.e. with constant M*, $\alpha_s$ and $\phi$*. In this heuristic case with a simple analytic solution (which we stress is *not* observed in the sky), the rate of cessation of star-formation is obtained by simply setting $dN_{blue}/dt = 0$ in equation (10) and inserting the expression (9) for $\alpha$ for a Schechter function:

$$\lambda + \kappa_- = -sSFR(\alpha + \beta)$$
$$= -sSFR\,(1 + \alpha_s + \beta - \frac{m}{M*}) \qquad (16)$$

There is now a degeneracy between the quenching of isolated galaxies and the merging rate *out* of the mass bin, $\kappa_-$, since clearly, in terms of depleting the star-forming population, there is no difference between cessation of star-formation in an isolated galaxy and cessation due to a merger. We will refer to the combination $\lambda + \kappa_-$ as the combined "death function", $\eta$, of star-forming galaxies, which is simply the probability that star-formation ceases in unit time, for whatever reason. The inverse $\eta^{-1}$ is a statistical timescale for quenching to occur (which should not be confused with the time required for the physical transformation to take place, once started).

It is instructive to consider the origin of the two terms on the right hand side of the second part of equation (16) in this heuristic example. The constant $1+\alpha_s+\beta$ term acts to keep the normalization of the Schechter function constant even though star-formation is bringing up a (generally larger) number of lower mass galaxies. The steepening of the star-forming mass-function that would occur if $\beta$ is not zero is counteracted directly by the small mass-dependence of the sSFR term. The most crucial point, however, is that the sSFR$\times m/M$* term acts to keep the characteristic M* in the Schechter function the same, as introduced above. As noted above, the $m$ dependence is required to produce the exponential cut-off at M*. The sSFR dependence comes because this controls the rate of logarithmic increase in stellar mass. Since the sSFR is simply the SFR/$m$, this term reduces to SFR/M*.

In the high mass regime around and above M*, equation (16) clearly reduces to the following very simple form for the (total) quenching rate (which will be dominated by mass-quenching), because $(1+\alpha_s+\beta)$ is small and $\kappa$ is small compared with $\lambda$:

$$\eta \sim \lambda_m \sim \frac{SFR}{M*} = \mu\,SFR \qquad (17)$$

Having established this at high masses, we can then look at what happens to the mass-function of star-forming galaxies if we were to postulate that the mass-quenching rate has this simple form (17) for galaxies of *all* masses, in *all* environments, and at *all* epochs.

Putting the expression for the slope $\alpha(m)$ of the Schechter function from equation (9) into the continuity equation (10), it can be seen that the mass-function of star-forming galaxies would drift up in normalization, unless $(1+\alpha_s+\beta)$ was zero (whereas it probably takes a value around $-0.5$). The rate of increase in $\phi$* will be proportional to the overall sSFR, and thus most of the evolution would occur at high redshifts where the sSFR is highest. Encouragingly, this is exactly what is seen in the data (Pérez-González et al., 2008, Pozzetti et al., 2009, Ilbert et al., 2010) as discussed in Section 3.4 above. In addition, adoption of this simple death function for all galaxies would imply that the faint end slope would gradually steepen with time, depending on the value of $\beta$ (with no change for the simple case of $\beta = 0$). As noted in Section 3.4, there is even evidence in the data for a small shift in this direction, especially at $z > 1$.

Since these changes to the star-forming mass-function are exactly what are observed, we are encouraged to postulate equation (17) to be the universal form of mass-quenching $\lambda_m$ for *all* galaxies. Although this very simple form of the mass-quenching is only *required* (by the constancy of M* of *star-forming* galaxies) at high masses, we show below (in Section 5.2) that it in fact reproduces the observed Schechter mass-function of *passive* galaxies over more than 2 orders of magnitude in mass, i.e. extending well below M* into the regime where the Schechter mass-function becomes a power-law. We will therefore conclude below that our postulated mass-quenching law (17) must be valid over this much broader (2 dex) range in mass, and will assume this for the remainder of this Section 4.

*4.4.4 The combined rate of the three quenching mechanisms*

We can now combine the "environment-quenching" and "merger-quenching" with the "mass-quenching" explored in Section 4.4.3. We demonstrated above that environment quenching must be independent of mass, because $\varepsilon_\rho$ is independent of mass (and time). Observationally, there is no clear indication of how the merging rate of galaxies depends on the galactic mass, and we will assume for simplicity that this is also independent of mass, like the environment quenching. Since both are the result of the growth of structure, this seems a reasonable assumption. On the other hand, mass-quenching will be roughly proportional to mass, *at any given epoch*, because of the dependence of SFR on mass.

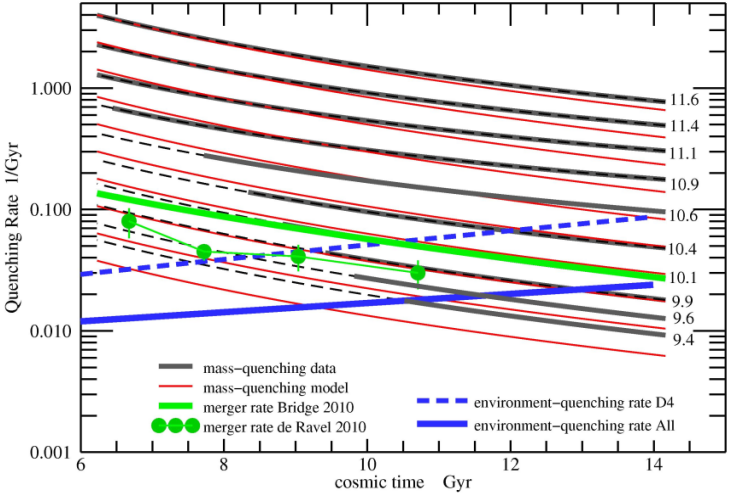

Fig 11: The mass-quenching rate $\lambda_m$ (in red and black), the environment-quenching rate $\lambda_\rho$ (in blue) and the merger-quenching rate $\kappa_-$ (in green) for galaxies as a function of cosmic time. The mass-quenching rates are shown for a number of different masses (log m shown on the right). These increase with mass because of the proportionality with SFR but fall with time due to the decline in cosmic sSFR. For the mass-quenching, the black curves (labeled as "data") are derived solely from observational quantities via equation (15), while the red curves (labeled as "model") are simply computed from equations (1) and (17). The (mass-independent) "environment-quenching" (in blue) is shown for the fourth density quartile D4, and for the median of all galaxies, using the fits from Fig 9. Environment-quenching increases with time as structure grows in the Universe. For the merger-quenching, two recent estimates of the merging rate are shown (in green). This is observed to fall with time and is assumed to be mass-independent. It should be noted that the blue and the green curves are both manifestations of the merging of dark matter haloes, depending on whether the incoming galaxies merge or survive as distinct entities, the latter occurring more at later times.

There is therefore a good motivation to consider a more general form for the combined quenching rate. We write the overall death function, $\eta$, acting on the population as a whole, as the sum of the star-formation driven term given in equation (17), plus the mass-independent term $\eta_\rho$ that is naturally associated with the environment-quenching rate $\lambda_\rho$, defined in equation (15), plus a (mass-independent) merging term $\kappa_-$. This may be written as

$$\eta = \lambda_m + (\lambda_\rho + \kappa_-)$$
$$= \mu SFR + \left(\frac{1}{1-\varepsilon_\rho}\frac{\partial \varepsilon_\rho}{\partial \log \rho}\frac{\partial \log \rho}{\partial t} + \kappa_-\right)$$
$$= \eta_m + \eta_\rho \quad (18)$$

Note that $\eta_\rho$ is independent of $m$ and $\eta_m$ is independent of $\rho$. At a given epoch, $\eta_m$ is proportional to $m^{(1+\beta)} \sim m$ for $\beta \sim 0$. Fig 11 shows the derived dependences of $\eta_m$, $\lambda_\rho$ and $\kappa_-$ on time and, for $\eta_m$ also on galactic stellar mass. The $\lambda_\rho$ and $\kappa_-$ are plotted for both the first and fourth environmental density quartiles, as derived above.

Comparing all of the rates, we conclude that mass-quenching dominates at all epochs for the highest mass galaxies, $m > 10^{10.5}$ M$_\odot$, i.e. at and above M*, while for low mass galaxies below $10^{10}$ M$_\odot$, environmental effects dominate at late epochs, with merging likely dominating at higher redshifts $z > 0.5$. We will return to this point below, but point out here the close connection between "merging" and "satellite-quenching". Both result from the merger of two dark matter haloes. If the baryonic galaxies merge, we will see a "merger". If they do not, we will see a new satellite galaxy which may be quenched through satellite-quenching.

### 4.4.5 The establishing of the Schechter function exponential cut-off

To see clearly the role of the quenching law (18) in establishing, as well as maintaining, the value of M*, we consider the change in the number of galaxies that is given by equation (10), together with the quenching law (18) applied to a Schechter function that initially has some different M*(t). We can then write:

$$\frac{1}{N_{blue}}\frac{dN_{blue}}{dt}\bigg|_m = -sSFR(\alpha + \beta) - (\lambda_m + \lambda_\rho + \kappa_-)$$
$$= -sSFR\left(1 + \alpha_s + \beta - m(M^*(t)^{-1} - \mu)\right) - \eta_\rho \quad (19)$$

Clearly if M*(t) is less than $\mu^{-1}$, the number of galaxies at high $m$ will increase strongly due to the term in the inner brackets. If M*(t) is greater than $\mu^{-1}$, the number of galaxies at high $m$ will likewise drop strongly. Only when M*(t) $= \mu^{-1}$ will the term in the inner brackets become zero and the change in the mass-function of star-forming galaxies be independent of mass (if $\beta = 0$). Thereafter, the evolution in the mass-function will be confined to changes in $\phi$* (with mild changes in $\alpha_s$ if $\beta \neq 0$).

In other words, if we start with some initial mass-function with generally low galactic masses, the mass-function of star-forming galaxies will initially traverse across to higher masses (with uniform shape if $\beta= 0$) due to star-formation. As significant numbers of galaxies approach $m \sim \mu^{-1}$, the quenching law acts to establish the exponential cut-off in the star-forming mass-function at M* $= \mu^{-1}$. At that point, and thereafter, a "garbage-pile" of passive galaxies will build up around this value of M* as star-formation continues to feed galaxies into the "buzz-saw" of the quenching. The overall mass-function will subsequently evolve only vertically in density, with fixed shape (if $\beta= 0$). The action of the mass-independent quenching terms, i.e. those due to environment- and merger-quenching, will be to cause additional passive galaxies to rain down from all masses.

To look at the evolution in $\phi$* once M* stabilizes, we simply integrate (19) with M*(t) = $\mu^{-1}$

$$\ln N_{blue}|_m = -\int_{t_{init}}^{t}\left(sSFR(1 + \alpha_s + \beta) + \eta_\rho\right)dt$$
$$N_{blue}|_m = N_{blue,init}|_m \left(\frac{m}{m_{init}}\right)^{-(1+\alpha_s+\beta)}\left(1 - f_{\eta_\rho}\right) \quad (20)$$

Here $f_{\eta_\rho}$ is just the fraction of galaxies that have suffered quenching (or merging) from the mass-independent $\eta_\rho$ term, and $m_{init}$ is the mass that galaxies seen at time $t$ would have had at time $t_{init}$, i.e. $m/m_{init}$ is the mass increase of these galaxies in the time interval of interest. If we substitute into equation (20) a Schechter mass-function at some initial time $t_{init}$, with a value of M* = $\mu^{-1}$ as in the death function (18), then we find that the mass-function remains, as expected, a Schechter function with exactly the same characteristic M*, a normalization $\phi$* that gradually *increases* with time by an amount that is related to the fractional increase in mass of star-forming galaxies, modulated by a *decrease* due to the mass-independent quenching due to the average $\eta_\rho$ in the population. Unless $\beta \sim 0$, the Schechter faint-end slope $\alpha_s$ will steepen, i.e. become slightly more negative, due to the mass-dependence of the $m/m_{init}$ term, which is of course independent of $m$ for $\beta = 0$. Both effects will be more prominent at high redshift where the sSFR is much higher and the mass growth term correspondingly larger. These results are of course the same as we argued above. If the initial M* is smaller than the $\mu^{-1}$ in our quenching-law, then the star-forming mass-function shifts to higher masses until M* reaches this value.

*4.4.6 Characteristics of mass-quenching*

We therefore argue that the observed evolution of the mass-function of star-forming galaxies, i.e. constant M* and a gradually increasing $\phi$* (Bell et al 2007, Pérez-González et al., 2008, Pozzetti et al., 2009, Ilbert et al., 2010) *demands* the remarkably simple form (17) for the mass-quenching of star-forming galaxies: one that is proportional only to the SFR and independent of stellar mass and epoch (except in so far as these two determine the SFR). To this must be added a mass-independent term, that will of course dominate at low masses, which can be identified by the environment-quenching plus mass-independent merging.

We should stress that although we have referred to a "mass-quenching" effect, because it produces the strong dependence of $f_{red}$ on galactic mass, the mass-quenching process is, paradoxically, *independent* of galactic mass per se. At a given epoch, the SFR of course depends strongly on mass since the sSFR has $\beta \sim 0$, but this is in our analysis a secondary correlation, although one which has an important effect on the shape of the mass-function, as discussed below. The primary dependence of quenching on SFR evidently holds rather well since $z \sim 2$, during which period the SFR, at a given stellar mass, has declined by a factor of 20.

We stress that the "M*" in our mass-quenching law equation (17) is a constant across the whole population of galaxies, and across a wide range of cosmic time. It therefore has nothing to do with the masses of *individual* galaxies, either in stars or in gas. Our mass-quenching law (17) does not therefore have much to do with the consumption timescales of gas in a particular galaxy or halo, and evidently has little to do with *any* properties of individual galaxies, *except* for their individual instantaneous star-formation rates.

The mass-quenching death rate is evidently simply the star-formation rate, divided by the (constant) characteristic mass M* of the Schechter mass-function of star-forming galaxies. The value of M* is empirically around $10^{10.6} M_\odot$, so from (17) we obtain quantitatively

$$\lambda_m \sim \left( \frac{SFR}{40 \, M_\odot yr^{-1}} \right) Gyr^{-1} \qquad (21)$$

Implicit in this scenario is a strong link between the star-formation rate and the star-forming lifetime of a galaxy. Applying equation (21) with star-formation rates of order 1000 $M_\odot yr^{-1}$, implies a lifetime $\eta_m^{-1} \sim 40$ million years, comparable to the implied lifetimes of the ULIRG phase (see e.g. Sanders & Mirabel 1996, Caputi et al., 2009). The massive ($m \sim 10^{10.7} M_\odot$) galaxies at $z = 2$ that are seen with high star-formation rates (SFR $\sim 100$ $M_\odot yr^{-1}$) must be close to the point of death, with quenching timescale $\tau_q \sim \eta_m^{-1} \sim 0.4$ Gyr, a few dynamical times at best. It will be interesting to see whether the characteristic dynamical properties of these objects (in-spiraling disk instabilities) that have recently been revealed by high resolution integral field spectroscopy (Genzel et al., 2008) are confined to these galaxies that are evidently on the point of being quenched, or are a more general phenomenon that is exhibited by less terminal galaxies at the same sSFR, but lower masses, at the same redshifts.

*4.4.7 Possible physical origins of mass-quenching*

The conclusion from the extended discussion in the previous sections is very simple: a quenching rate that is simply proportional to the SFR establishes and maintains a Schechter mass-function for star-forming galaxies with a value of M* that is constant, despite the increase in individual stellar masses.

As we have stressed, our analysis is purely empirical and there may or may not be a direct causal link with star-formation. As a trivial example, if the mass-quenching rate was for some reason proportional to galactic mass and to some other quantity that was proportional to the cosmic epoch to the inverse 2.2 power, it would look exactly as if it was proportional to the SFR, even if there was no direct physical link with star-formation. We explore this aspect further in Section 7.3 below, where we show the links between our scenario, involving linking the quenching *rates* and star-formation *rates*, and a picture involving a static "mass limit" for galaxies.

This statistical analysis cannot establish such a causal link. Rather the goal has been to identify the signatures of the evolution that any more physically based model must obey. Having said that, what physical process could cause a quenching rate that is indeed causally proportional to star-formation rate, independent of stellar mass and epoch? One obvious possibility is some feedback process involving energy-injection from supernovae and the creation of superwinds, or some other process such as the local ionization model of Cantalupo (2010). Nevertheless, it is surprising that the depth of the potential well, or some other properties of the local environment, do not appear to enter.

Another possibility could be related to AGN-feedback. In Silverman et al. (2009b) (see also Daddi et al., 2007b) we showed that the ratio of the black-hole accretion rate $dm_{BH}/dt$ and the star-formation rate in zCOSMOS X-ray selected AGN was roughly constant, at least back to $z \sim 1$, with a mass ratio of order 0.01. A similar ratio was found in the SDSS by Netzer et al (2009). Multiplying this ratio by the observed AGN-fraction of about 0.03 for these same objects (Bundy et al., 2008, Silverman et al., 2009a) gives an average black-hole accretion rate of 0.0003 SFR. This means that the chance of having a quasar-like outburst could be (statistically) proportional to the SFR. Equation 21 could therefore be re-written (at least statistically) in terms of a black-hole accretion rate, BHAR.

$$\eta_m = 80 \left( \frac{BHAR}{M_\odot yr^{-1}} \right) Gyr^{-1} \qquad (22)$$

We note that if this connection with AGN is valid, this would imply that powerful quasars with BHAR $\sim 1$ $M_\odot yr^{-1}$ should lead to rapid quenching with $\tau_q \sim 10^7$ yr.

Regardless of the actual physical mechanism, our model can be used to make a clear prediction for the statistical properties

(i.e. their mass-function) of those objects that are being seen in the *process* of being quenched. This prediction, which may help in identifying plausible transient objects, and thus lead to a physical understanding of mass-quenching, is developed in Section 5.6 below.

## 5. THE ORIGIN OF THE SCHECHTER FUNCTION(S) AND THE VALUE OF THE CHARACTERISTIC MASS M*

In the previous Section 4.4.3 we argued that, from a purely observational stand-point, the quenching of star-forming galaxies must follow, at least at high masses, a very simple law, in which the galaxy death rate is given statistically only by the star-formation rate, plus an additional mass-independent term that accounts for the environmental-quenching, driven by the slow growth of structure over cosmic time, plus any mass-independent merging. We stress that this empirical conclusion regarding mass-quenching is demanded by the exponential cut-off in the Schechter shape of the mass-function of *star-forming* galaxies and the observed constancy of M* over a wide range of epochs since z = 2 coupled with the uniformity of the sSFR with mass. It is also completely consistent with the slow increase in φ*.

In this Section, we examine the role of the three processes of mass quenching, environment quenching and merger quenching in setting up also the mass-function of *passive* galaxies. We shall find that the application of the star-formation "law" down to low masses naturally accounts for the observed mass function of the passive galaxies, and this in itself may be taken as establishing the validity of this postulated extension.

### 5.1 The origin of the single Schechter mass-function of star-forming galaxies

We showed in Section 4.4.3 how the observed constancy of M* of star-forming galaxies requires a particular form of the mass-quenching rate, i.e. one proportional to the star-formation rate alone. We can turn the argument around and suggest that a mass-quenching rate for star-forming galaxies that is proportional simply to the instantaneous star-formation rate could be *the* physical mechanism for producing the distinctive Schechter function for star-forming galaxies, and the observed constant characteristic mass M*, independent of whether and how the actual sSFR changes with time. This idea is supported by the analysis in Section 4.4.5 which shows how this quenching law can also establish the exponential cutoff at this particular mass, as well as simply maintaining it.

In this picture, there is thus a very simple relation between the constant of proportionality between the star-formation rate and the mass-quenching death rate, m, and the resulting characteristic mass of the Schechter function. If the mass quenching death rate is given by

$$\eta_m = \mu \text{SFR} \quad (23)$$

with µ some constant reflecting the physics of the quenching process, then the result will be to produce a Schechter mass function for the star-forming galaxies, with a characteristic mass M* given by:

$$M^* = \mu^{-1} \quad (24)$$

Although we cannot identify the value of µ in physical terms, we suggest that it is this parameter that actually determines M* (and not vice versa).

We established in Section 4.4.3 that the action of our mass-quenching law is to produce a single Schechter function for the star-forming galaxies with this characteristic M*. Any additional quenching that is *independent* of mass will cull galaxies uniformly from the mass-function, and will not change its shape.

### 5.2 The origin of the two-component Schechter mass-function of passive galaxies

Quenching will halt further mass growth, except through the effects of merging, and so the mass-function of *passive* galaxies should be closely related to that of the blue *star-forming* galaxies which in turn, we have argued, should be determined by the mass-quenching parameter $\mu$ in equation (18).

At any epoch, the star-formation rate (and thus quenching rate) in star-forming galaxies will be proportional to $m^{1+\beta}$. Thus, if the star-forming population at some epoch obeys a Schechter function with some $M^*_{blue}$ and $\alpha_{s,blue}$, then the passive galaxies produced (per unit time) by mass-quenching of this population will have the same mass-function shape, multiplied by $m^{1+\beta}$. Given the form of the Schechter function,

$$\phi(m)dm = \phi^* \left(\frac{m}{M^*}\right)^\alpha e^{-m/M^*} \frac{dm}{M^*} \quad (25)$$

the passive mass-function will therefore be a new Schechter function with *exactly* the same value of M*, but with a faint-end slope $\alpha_{s,red}$ that is modified by (1+β), i.e.

$$\alpha_{s,red} = \alpha_{s,blue} + (1+\beta) \quad (26)$$

This will be true at a given epoch and true at *all* epochs if the $\alpha_s$ and M* of the star-forming galaxies are unchanging, as is more or less observed, as in Sections 3.4 and Table 1.

We will therefore get a new Schechter function with exactly the same M* but with a modified faint end slope which will differ by an amount that is related to the slope of the sSFR-*m* relation, $\Delta\alpha_s = (1+\beta) \sim 0.9$, or $\Delta\alpha_s = 1$ for the simple case of β = 0. This is a clear consequence of the simple mass-quenching law that we have adopted, *provided that it extends down to masses well below M*,* as we have postulated.

It has been known for many years that the Schechter function for (massive) passive galaxies has a faint end slope that is substantially less negative than that of star-forming galaxies (e.g. Binggeli, Sandage and Tammann 1988, Lilly et al., 1995, Folkes et al. 1999, Blanton et al. 2001, Ball et al. 2006). Already a decade ago (Folkes et al., 1999), it was established that the faint end slope differs from that of star-forming galaxies by $\Delta\alpha \sim 1$. Our quenching law, derived (and extended) from the time-dependence of the mass-function of *star-forming* galaxies at high masses, naturally produces this striking feature of the *passive* galaxy population over a much broader range of masses.

The result of the mass-*independent* term(s) in the overall death function given by equation (18), which we associated with the environmental quenching and/or mass-independent merging of galaxies, will be to produce a second population of passive galaxies whose mass-function simply shadows the mass-function of star-forming galaxies, with exactly the same M* and $\alpha_s$, but (generally) lower $\phi$*. This second Schechter component will therefore dominate the mass-function of passive galaxies at low masses.

There is good evidence for the existence of this second component of passive galaxies. An upturn in the mass-function of passive galaxies at low masses is clearly seen in the SDSS *r*-band luminosity function of Blanton et al. (2001) (see also our own SDSS analysis below) and has also been highlighted in the deep photo-*z* based COSMOS analysis of Drory et al. (2009, see their Fig 5). We would expect the size (i.e. $\phi$*) of this second, lower mass, population of passive galaxies, to be larger in higher density environments, because of larger $\lambda_p$ term in equation (18), the higher merging rate, and the requirement that $f_{red}$ increase with density for all masses, including these very low ones. All other things being equal, we predict that the strength of the second (low mass) component would be a few (approximately four) times stronger in the highest density quartile than in the lowest quartile, reflecting the higher environment quenching and merging rates in the denser environments.

To summarize: if we neglect for the moment the effects of any subsequent merging of quenched galaxies (see Section 5.5 below), then we would expect to see a double (two-component) Schechter function for the passive galaxies. The two components will have exactly the same M* but have two $\alpha_s$ that differ by (1+β), one of which will be the same as that of star-forming galaxies and the other shallower (i.e. less negative) by an amount related to the slope of the sSFR-*m* relation. Furthermore, the common M* value should precisely match that of the star-forming galaxies.

The two Schechter components in the mass-function of passive galaxies therefore reflect the two distinct quenching routes. At any given epoch, one is roughly *proportional* to mass, because of the flat sSFR-mass relation, and the other is quite *independent* of mass, through the separability of mass and environment established in this paper.

*5.3 The origin of the two-component Schechter function for all galaxies.*

If we now add together the mass-functions of both the star-forming and passive galaxies, we would therefore simply predict another double (i.e. two-component) Schechter function. Both components will again share the same M* with the two $\alpha_s$ differing as above by 1+β ~ 0.9. This is again a simple prediction of the adopted death function given by equation (18).

*5.4 Modification of the Schechter function(s) by subsequent merging.*

The above discussion considered the effect of merging in the *quenching* of galaxies, but did not include the changes in the mass-functions due to either the increase in the masses and reduction in the number-densities of the merger remnants. These two effects can be thought of as translating some fraction of the mass-function to lower number densities and higher masses. For galaxies at low masses, which were generally *quenched* by merging or by our environment-quenching process, and where which the mass-function is a power-law, these two effects will not substantially change the *shape* of the mass-function.

However, for galaxies which were originally *mass-quenched*, i.e. those around M* where the mass-function is strongly peaked (with $\alpha_s$ ~ –0.4) the effect of subsequent merging on the mass-function may be significant, especially at the highest masses where the mass-function steepens dramatically. These galaxies also generally quenched early, and therefore have more time for subsequent merging. Most of the mergers at these masses will be "dry", i.e. involve already-quenched galaxies, simply because these dominate the galaxy population. If the merger rate is independent of mass (and mass-ratio), this subsequent dry-merging will produce shadow mass-functions, that are simply displaced to higher masses and lower densities. These will likely dominate the resulting total mass-function at very high masses where the mass-function is steep.

The resulting mass-function of passive galaxies will therefore be more complex than the single Schechter function(s) described above. However, if the composite mass-function (summing the non-merged and the displaced merged populations) is force-fit by a single Schechter function, then the result is generally to increase M* and decrease $\alpha_s$ (i.e. make it more negative). To illustrate this effect, we construct a heuristic composite mass-function in which we assume that 15% of galaxies have undergone a 1:1 merger. This is done by adding a 7.5% component of galaxies with $\Delta m$ = +0.3 dex to an 85% unmerged population that is described by a single Schechter function with $\alpha_s$ = -0.4. We then force-fit this new composite population with a single Schechter function. In this particular case, $\alpha_s$ decreases by 0.15 and M* increases by 0.09 dex, i.e. $\Delta\alpha_s$ ~ 1.6 $\Delta$M* for small amounts of merging.

Interestingly, a lot of this shift in M* is associated with the degeneracy between M* and $\alpha_s$ in Schechter fits, since the *average* mass of the galaxies in this simple illustration has evidently increased by only 0.03 dex, i.e. only *a third* of the change seen in M*. Merging at larger mass ratios would require a larger fraction of mergers to produce the same distortion of the mass-function and would produce a shift in M* that was closer to the change in average mass.

Provided that the amount of post-quenching merging of passive galaxies is modest, it evidently produces a relatively small perturbation to the predictions described in the previous section. It should be noted that even though the merger rate for passive galaxies is overall small, the most massive galaxies on the steepest part of the mass-function will almost all have undergone significant merging, as we discuss further in Section 6 (see Fig 16 below).

*5.5 Observational tests of the predicted relations in the SDSS mass-functions.*

The above analysis made five simple predictions for the forms and inter-relationships of the mass-functions of star-forming and passive galaxies in different environments in our simple empirically driven model:

(1) The mass-function of star-forming galaxies should be a pure single Schechter function, but that of passive galaxies should be a double Schechter function.

(2) The mass-functions of star-forming galaxies in different environments should quantitatively have the same values of M* and $\alpha_s$.

(3) The mass-functions of star-forming galaxies and of the passive galaxies in the lowest density D1 quartile (where post-quenching merging can be neglected) should have exactly the same M*, but values of as that differ by (1+β) ~ 1.

(4) The secondary component of passive galaxies, which dominates at the lowest masses, should have the same M* and α as the star-forming galaxies, and (easier to verify) an amplitude

Table 3: Schechter function parameters for SDSS mass-functions

| Sample | Log(M*/M☉)[a] | φ₁*/10⁻³Mpc⁻³ | α₁ | φ₂*/ς | α₂ |
|---|---|---|---|---|---|
| *(a) Free fitting parameters* | | | | | |
| Global | 10.67 ± 0.01 | 4.032 ± 0.12 | -0.52 ± 0.04 | 0.655 ± 0.09 | -1.56 ± 0.12 |
| Blue-all | 10.63 ± 0.01 | ... | ... | 1.068 ± 0.03 | -1.40 ± 0.01 |
| Blue-D1 | 10.60 ± 0.01 | ... | ... | 0.417 ± 0.02 | -1.39 ± 0.02 |
| Blue-D4 | 10.64 ± 0.02 | ... | ... | 0.151 ± 0.01 | -1.41 ± 0.04 |
| Red-all | 10.68 ± 0.01 | 3.410 ± 0.07 | -0.39 ± 0.03 | 0.126 ± 0.02 | (-1.56) |
| Red-D1 | 10.61 ± 0.01 | 0.893 ± 0.03 | -0.36 ± 0.05 | 0.014 ± 0.01 | (-1.56) |
| Red-D4 | 10.76 ± 0.02 | 0.814 ± 0.03 | -0.55 ± 0.06 | 0.052 ± 0.01 | (-1.56) |
| *(b) With fixed α₁, α₂, and M*, varying only φ** | | | | | |
| Blue-all | (10.67) | ... | ... | 1.014 ± 0.02 | (-1.40) |
| Blue-D1 | (10.67) | ... | ... | 0.369 ± 0.01 | (-1.40) |
| Blue-D4 | (10.67) | ... | ... | 0.149 ± 0.01 | (-1.40) |
| Red-all | (10.67) | 3.247 ± 0.07 | (-0.4) | 0.214 ± 0.02 | (-1.40) |
| Red-D1 | (10.67) | 0.812 ± 0.03 | (-0.4) | 0.023 ± 0.01 | (-1.40) |
| Red-D4 | (10.67) | 0.864 ± 0.04 | (-0.4) | 0.111 ± 0.02 | (-1.40) |

Note. [a]) The M* for both Schechter components are assumed to be the same.

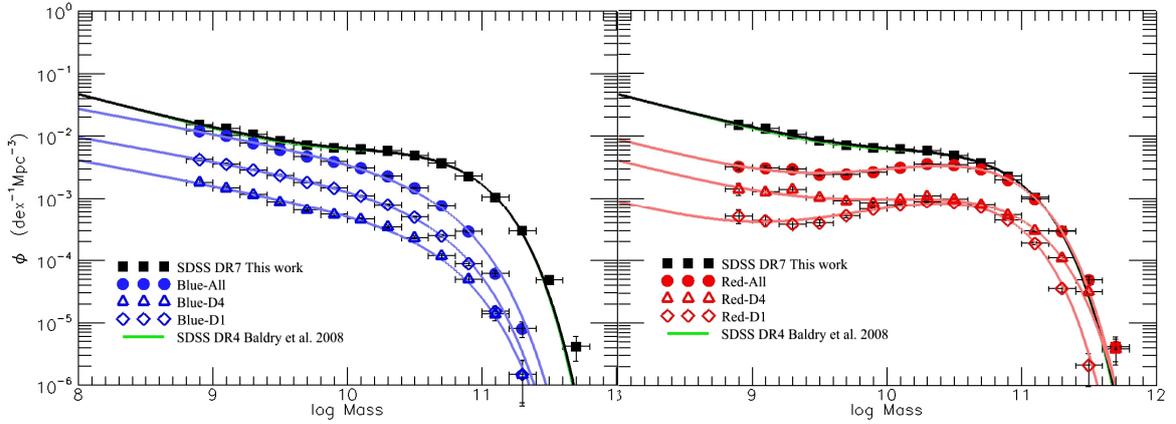

Fig 12: The mass-function of SDSS galaxies split into blue and red galaxies (left and right panels) and by environmental (within each panel). The total mass-function (i.e. all galaxies and all environments) is shown in both panels as a black line, overplotted against a green line which is the equivalent total mass-function from Baldry et al. (2008). These two are almost identical. The horizontal bars indicate the size of the mass bins, the vertical bars indicate the statistical uncertainties.

φ* that is higher (by about a factor of 4) in the high density D4 environment compared with D1.

(5) The primary component of the mass-function of passive galaxies should have been modified in D4 through post-quenching merging, producing a higher M* and less negative $\alpha_s$, with $\Delta\alpha_s \sim 1.6 \Delta M*$.

We have tested the first five simple predictions by constructing the mass-functions of the SDSS red and blue samples, plus the overall mass-function, in the highest (D4) and lowest (D1) quartiles of environmental density, using the same $V_{max}$ approach that we described above. These are shown in Table 3 and Fig 12. We also show on Fig 12 the overall mass-function constructed from the SDSS DR4 by Baldry et al. (2008), which shows excellent agreement with our own overall mass-function.

For the blue galaxies we use only a single Schechter function since this provides a fully acceptable fit, while for the red galaxies we use a double (two-component) Schechter function, with the M* of the two components constrained to always have the same value, it being almost impossible to constrain the M* of the weaker component independently. As is clear from Fig 12, the mass-function of passive galaxies indeed requires a two-component Schechter function. We also find we have to fix the $\alpha_s$ of the weaker of the two passive components to be the faint end slope of the blue population since this also is poorly constrained by the data.

All of the basic quantitative relationships predicted by our simple model are seen to hold rather well. For example, the M* for passive and star-forming galaxies in the overall mass-function (not differentiating at all by environment) differs by only 0.05 ± 0.02 dex.

Turning to the lowest density D1 quartile, where we could expect the effects of merging to be minimal, we find that the two M* (blue and red) are essentially identical, i.e. differing by 0.01 ± 0.02 dex, i.e. well within the observational uncertainties. The $\alpha_s$ of the dominant population of passive galaxies has $\Delta\alpha_s = 1.03 \pm 0.05$ relative to the blue galaxies, compared with the prediction of $\Delta\alpha_s = (1+\beta) = 0.9$ for $\beta = -0.1$, and of course $\Delta\alpha_s = 1.0$ for the simple case of $\beta = 0$. We take these results as a compelling vindication of our simple model.

Looking at the overall mass-function for all galaxies, we find $\Delta\alpha_s = 1.04 \pm 0.13$ for the two Schechter components (with M* constrained to the same). In their analysis of the SDSS DR4, Baldry et al. (2008) found also a double Schechter function with $\alpha_s$ of -1.58 and -0.46, i.e. with $\Delta\alpha_s \sim 1.12$, i.e. consistent with our own analysis, and also with the expected value of $\Delta\alpha_s$ from our model.

Finally, turning to the highest density quartile, D4, it can be seen that the $\alpha_s$ of the dominant passive population is a little more negative, by $\Delta\alpha_s \sim 0.2 \pm 0.08$, in the densest D4 quartile compared with D1, and that M* is also bigger by about 0.16 ± 0.02 dex. As discussed in Section 5.4, these small differences can be easily explained by a modest amount (≤ 30%, and possibly < 15%) of dry-merging of the passive galaxies in the D4 environment (and none in D1). In contrast, the *blue* population in the different environments are very similar, i.e. the two M* are within 0.04 ± 0.03 dex and the two $\alpha_s$ are within 0.02 ± 0.04. This compelling similarity is again expected in our simple model, since the shape of the mass-function for the blue galaxies will not be at all affected by merging, provided that the merger rate is independent of mass and that mergers remove galaxies from the blue main sequence.

We draw from the above three important conclusions:

(1) our simple model works extremely well, especially in the low density D1 quartile where the effect of merging around M* is negligible;

(2) the success in correctly predicting the mass-function of *passive* galaxies over a broad range of masses extending well below M* shows that our mass-quenching law must be valid over a much broader range of masses (of *star-forming* galaxies) than it was originally derived from (see Section 4.4.3); and,

(3) the amount of merging in even dense environments is quite constrained, if our model is correct. We would be very surprised, given the observed increase in M* of 0.16 dex, if the average mass of typical passive galaxies (around M*) in the densest quartile had increased by more than this same amount (i.e. 40%), and, in fact, given the factor of 3 gain in ΔM* relative to the mean increase in galactic stellar mass discussed in Section 5.4, we suspect the change in the average mass of these galaxies due to dry-merging is probably considerably smaller, i.e. possibly as low as 0.06 dex (15%).

Given that passive galaxies clearly dominate the high mass end of the mass-function (e.g. Figs 4 and 12), readers may be surprised by our conclusion that the blue sequence of star-forming galaxies can populate the red sequence with little or no post-quenching dry-merging, completely so in D1, and with only 15-40% merging in D4 (c.f. the different scenarios of Faber et al., 2007). We stress that this conclusion follows (if our model is correct), from the observation that the M* of active and passive galaxies are essentially identical. Although passive galaxies clearly vastly outnumber star-forming ones at the highest masses, the most massive star-forming galaxies are so rapidly quenched that they are visible for only a short period of time, making it fully possible for the red sequence to be populated directly from the blue sequence, at least in D1.

It is often thought that the color distribution of galaxies as a function of mass (e.g. Fig 4) indicates a sharp mass "threshold" between star-forming (lower mass) and passive (higher mass) galaxies and there have been speculations as to the evolution of this transition mass. We stress however that red and blue galaxies in fact form a *completely overlapping* population with essentially the same M*. In this context, it should be appreciated that two Schechter functions that have the same M* but very different $\alpha_s$ (as in the blue and red populations) "look" quite different, even around M*. As an example, the "typical" mass of galaxies, i.e. the mass at which $m\phi(m)$ peaks, differs by 0.6 dex for two Schechter functions with the same M* but with $\Delta\alpha_s = 1$. Likewise, the relative numbers of galaxies in the two populations shift from one population to the other as $m^{\Delta\alpha}$.

To gain an appreciation of the relative sizes of the different components of the galaxy population in different environments, we fix all of the Schechter parameters except $\phi^*$ (to avoid the well-known degeneracies). These are shown in the second part of Table 3. The values of $\phi^*$ in each quartile of density are arbitrary, since they depend on the definition of the quartile. The most useful approach is to normalize each component to the $\phi^*$ value for the dominant ($\alpha_s = -0.4$) passive population in the two density quartiles. This yields relative normalizations for the two $\alpha_s = -1.4$ components of 0.45 (blue D1), 0.17 (blue D4), 0.028 (red D1), and 0.13 (red D4).

Normalizing in this way, we find that the secondary passive component is some 4.5 times larger in D4 compared with in D1. This reflects the higher environment-quenching and merger rates in the denser environment building up this component.

It can also be seen that the sum of the two $\alpha_s = -1.4$ components (blue + red) is 60% larger in D1 (0.48) than it is in D4 (0.30), when both are normalized to the respective $\alpha_s = -0.4$ passive populations. This is the effect we drew attention to in Bolzonella et al. (2009), in which the "hump" of the double Schechter function is more pronounced in the higher density environments. We suspect that it reflects a difference in the mass-function of star-forming galaxies at very early times in the denser environments which leads to more galaxies reaching the mass-quenching regime around M* before the sSFR starts to drop. This is illustrated in our numerical model in Section 6.

Finally, although they are most beautifully established in SDSS, we note that there is quite good evidence in favor of these basic relationships at higher redshifts also. If we average the fitted mass-function parameters in zCOSMOS for 0.1 < z < 0.7, from Bolzonella et al. (2009), which mostly samples relatively massive galaxies, we find the same values of M*, with log M*/M☉ = 10.77 and 10.82 for red and blue galaxies respectively, and with $\alpha_s$ = -0.32 and -1.4, i.e. a $\Delta\alpha_s$ slope difference of 1.1, close to our prediction of about 1.0.

*5.6 The mass-function of galaxies in the process of being mass-quenched*

Regardless of the physical origin of mass-quenching, there is a clear prediction from our model for the mass-function of those galaxies that are in the *process* of being mass-quenched. This prediction may help in identifying plausible candidates for such objects, be they associated with AGN or some other transient phenomenon such as unusual color-morphology combinations, etc.

We assume that the process of mass-quenching is associated with some recognizable transient signature that is visible for some period of time $\tau_{trans}$. The simplest assumption would be that $\tau_{trans}$ is independent of our variables of $m$, $\rho$ and $t$. In this case, it is easy to see from equation (10), and setting $\kappa = 0$, that the mass function of the transient "being-quenched" objects should follow:

$$\phi_{trans}(m,t) = \frac{dN_{red}}{dt} \tau_{trans}$$
$$= \phi_{blue}(m,t) \times SFR(m,t) \times \mu \, \tau_{trans}$$
$$= \phi_{blue}(m,t) \times sSFR(m,t) \times \left(\frac{m}{M^*}\right) \tau_{trans} \quad (27)$$

Therefore, the mass-function of the transient objects will be given, for the simple case of $\tau_{trans}$ = constant, by a single Schechter function with parameters

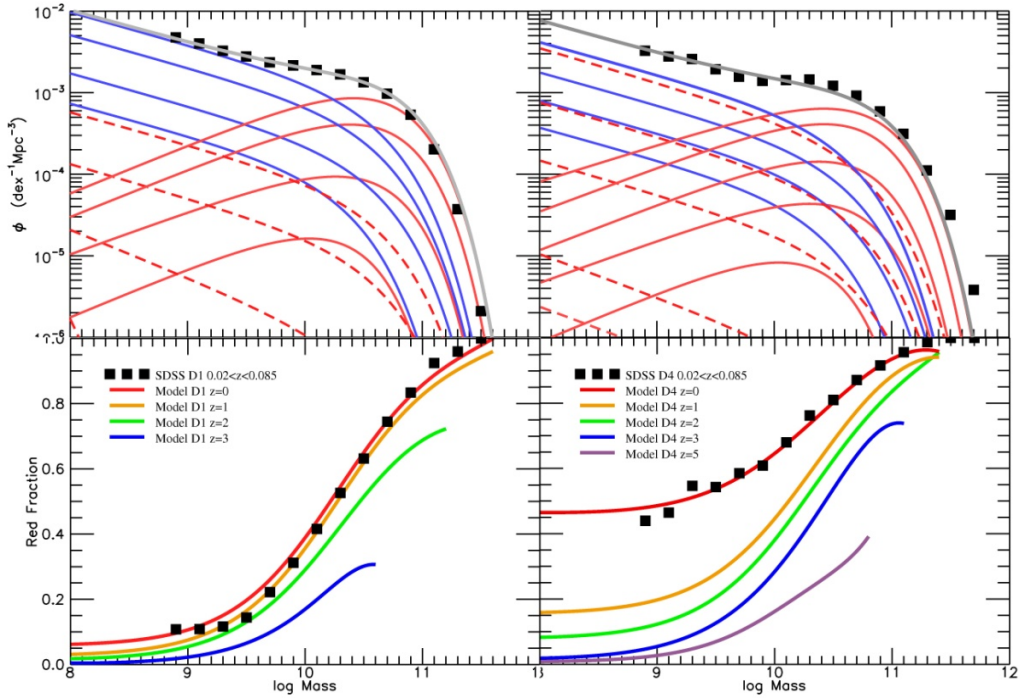

Fig 13: Evolving mass-functions (top) and red fractions (bottom) for the simple model described in the text. The solid blue lines indicate the mass-function of star-forming galaxies, the red lines represent the "mass-quenched" passive galaxies, and the dashed red lines show the "environment-quenched" and "merger-quenched" passive galaxies. The model is computed for the lowest density D1 quartile (left, shown for z =3, 2, 1, 0) and for the highest density D4 quartile (right, at z = 5, 3, 2, 1, 0, although the latter two blue mass-functions are completely overlapping). Also shown in all panels are the low redshift observational data from the SDSS survey. The black line in the left-hand panel shows also the model prediction and SDSS data for all galaxies (regardless of environment). In each environment, the model is normalized in total number to the SDSS data. A movie based on this figure is available in the online journal.

$$M^*_{trans} = M^*_{blue} = M^*_{red}$$
$$\alpha_{s,trans} = \alpha_{s,blue} + (1+\beta) = \alpha_{s,red}$$
$$\phi^*_{trans} = \phi^*_{blue}\, sSFR(t)|_{M^*}\, \tau_{trans} = \phi^*_{blue}\, \frac{\tau_{trans}}{\tau(t)|_{M^*}} \quad (28)$$

where $sSFR(t)|_{M^*}$ and $\tau(t)|_{M^*}$ are the evolving sSFR and star-formation timescale respectively ($\tau = sSFR^{-1}$), evaluated at the galactic mass corresponding to M*.

The *shape* of the transient mass-function should therefore be exactly the same as for the population of *already* mass-quenched *passive* galaxies, but the number density normalization $\phi_{trans}*$ will be the product of the $\phi_{blue}*$ of the *currently star-forming* galaxies multiplied by the dimensionless ratio of the visibility time-scale $\tau_{trans}$ and the star-formation timescale $\tau$ at M. This ratio will strongly evolve with redshift due to the change in sSFR given by equation (1). More general cases with a non-constant $\tau_{trans}(m, t)$ may be easily derived.

As would be expected, this formalism naturally produces a close linkage between the evolution of the characteristic sSFR in the Universe as a whole and the number density of the transient objects involved in the mass-quenching process.

An important consequence of our model is that all of these statements regarding the mass-function of transitory objects should be strictly *independent* of the Mpc-scale environment. This is simply because of the independence of mass-quenching on the environment that was established in Section 4.2.

Having established these inter-connections, we could ask whether the transient objects are actually seen to be still on the star-forming main-sequence, or are already passive galaxies, or intermediate objects, or some combination of all three. This information would tell us when in the mass-quenching process the relevant "signature" was produced, but the fundamental relationships in the mass-functions that we have outlined above would be unchanged.

## 6 A SIMPLE MODEL FOR THE EVOLVING GALAXY POPULATION

We believe that the extremely simple picture described above is a very attractive empirical model for the broad evolution of the galaxy population, at least since $z \sim 2$. In fact, since nothing in it is redshift dependent, we suspect that it is also likely to be applicable at very much higher redshifts. Of course, some aspects may be different at the highest redshifts. For instance, it may be that merging at very high redshifts does not lead to quenching. Correspondingly, it may be that the revival of quenched galaxies through gas-replenishment is much more common at earlier times.

The development of the model was based on a number of very simple features of the galaxy population at low and high redshifts. To illustrate the model with a concrete example, we have constructed a very simple simulation of the evolving galaxy population over the whole redshift range. This has the following inputs.

We take the death function at all times and for all galaxies to be as in equation (18). We use a value of $\log (M^*/M_\odot) = 10.6$, and use the curve for $\varepsilon_\rho(\rho)$ from Fig 8, which we assume to be completely independent of epoch. We take the mean $d\log\rho/dt$ for Quartile 4 from Fig 9, extrapolating beyond $z = 1$ where the effects of environment will anyway be rather small. For the lowest density D1 quartile we take $d\log\rho/dt \sim 0$ (see Fig 9). We use equation (1) as the sSFR history for all non-quenched galaxies, and adopt $\beta = 0$ for simplicity, i.e. a sSFR independent of mass. Finally, we take the merging rate $\kappa_-$ as $0.027(1+z)^{1.2}$ in D4 quartile, and four times lower in D1 (taken from de Ravel et al., in preparation, see also Kampczyk et al., in preparation). This merger rate is extrapolated at $z > 1$, but merger-quenching is turned off at $z > 3$, although this does not have a major effect.

For each of the two quartiles D1 and D4, we generate a sample of 6 million star-forming galaxies at $z = 10$ (i.e. 0.5 Gyr after the Big Bang), following a primordial logarithmic mass-function that is a power-law with slope equivalent to a Schechter faint end slope of $\alpha_s = -1.4$, plus an arbitrary cut-off at high mass. This cut-off is simply required to avoid over-populating the initial population with extremely massive galaxies, but the precise value is immaterial (because of the way our quenching law automatically sets up the correct value of M*, independent of starting point). In order to reproduce the effect described in Bolzonella et al. (2009), i.e. the relative strengths of the components with $\alpha_s \sim -0.4$ and $\alpha_s \sim -1.4$ populations in different environments, we start the simulation 1 Gyr later in the D1 quartile. Given the e-fold time for galaxy mass in the exponential growth phase, this is equivalent also to increasing the masses of galaxies in D4 by a factor $e^2 = 7.4$ and starting the simulation at the same epoch.

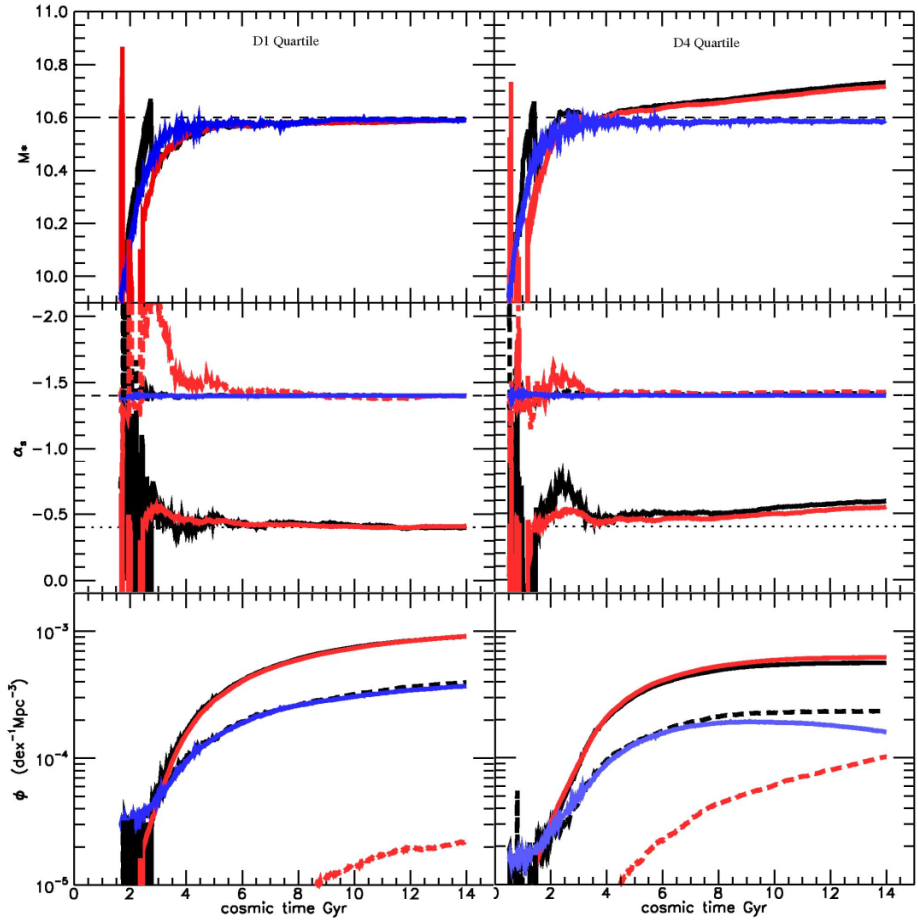

Fig 14: Evolution over cosmic time of the Schechter parameters for different components of the galaxy population in the simple simulation described in the text. The lines are as in Fig 13. The rapid establishment of the various Schechter functions is clear. In D1, where the effects of merging are minimal, the values of $\alpha_s$ and M* quickly stabilize and there is a gradual increase in $\phi$*. In D4, post-quenching dry-merging acts to cause both M* and $\alpha_s$ of the dominant component of the passive galaxies to slowly evolve to brighter M* and more negative $\alpha_s$, as described in the text.

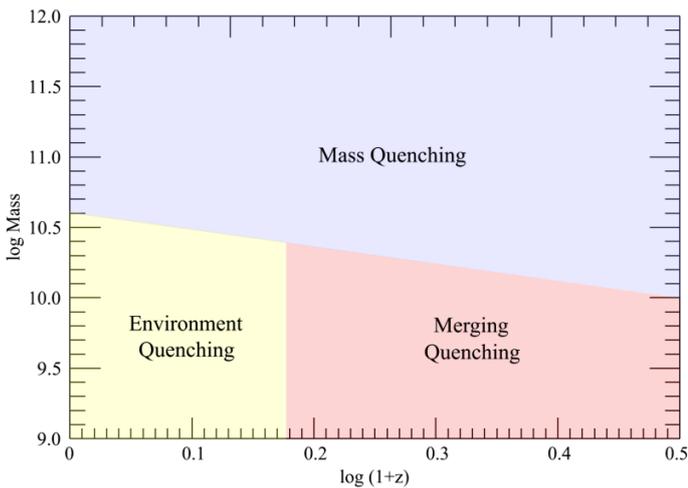

Fig 15: Diagram showing the dominant mechanism for the quenching of galaxies as a function of mass and redshift in typical (median) environments. As discussed in the text, merging and environment-quenching, although very distinct observationally and having very different consequences for the galaxies involved, both reflect the underlying merger of dark matter haloes.

The galaxies then evolve according to the above relations. Figure 13 shows the evolving mass-function in the D4 and D1 density quartiles for star-forming and passive galaxies, at a few representative redshifts, plus the overall mass-function at the present epoch, $z = 0$. The lower panel shows the red fraction of galaxies at different masses at these same redshifts. Figure 14 shows the evolution of the $\phi$* for red (both Schechter components), blue, and all, galaxies with cosmic time. A movie showing the continuous development of these relations is published in the electronic journal.

Based on the model, but reflecting the underlying discussion of the transformation rates associated with Fig 11 above, we show in Fig 15 the dominant mode by which star-formation ceased in passive galaxies as a function of their mass and redshift at the time of their quenching. Above a certain mass, which slowly increases with time, mass-quenching (i.e. that associated simply with the star-formation rate) dominates. Below this mass, merging and the environmental effect that we have associated with satellite quenching dominate. Merging dominates at earlier times, but satellite quenching comes to dominate at relatively late epochs, $z < 0.5$. As noted above, both merging and satellite quenching are clearly different manifestations of the underlying merger of two dark matter haloes. Although they produce observationally very different signatures and represent, of course, catastrophically different outcomes for the galaxies themselves, they are, from a dark matter point of view, associated with the same process of the hierarchical assembly of dark matter haloes.

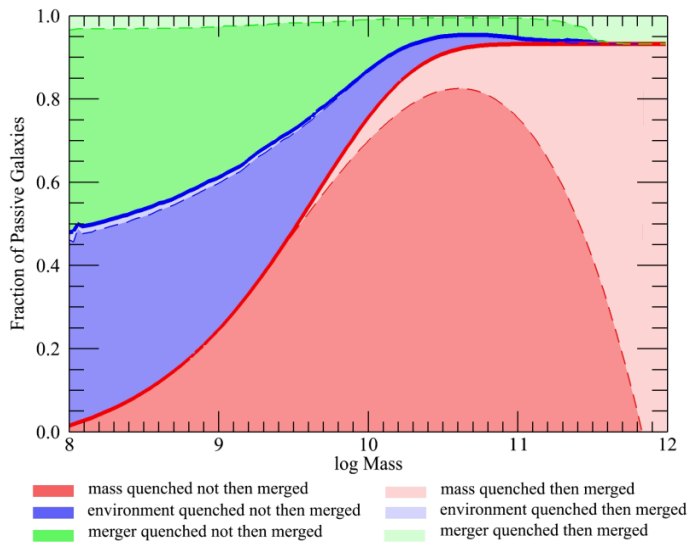

Fig 16: Diagram summarizing the evolutionary histories of today's passive galaxies (summed over all environments) as a function of their final stellar mass. The colors represent different modes by which the passive galaxies were initially quenched, i.e. mass quenching (red), environment quenching (blue) and merger quenching (green). The color shades then represent whether the galaxy subsequently underwent a merger (yes: light, or no: deep). The effect of post-quenching merging on environment-quenched galaxies is small, because most environment-quenching takes place after the merging rate has declined (see Fig 15). Although the amount of post-quenching merging of mass-quenched galaxies is quite small (only a few % in the overall galaxy population), and although the rate of merging is assumed to be independent of mass, the steepness of the mass-function above M* means that dry-merging will have been progressively more important in the most massive galaxies above $10^{11}$ M$_\odot$.

Although the effects of dry-merging are on average small for the population of passive galaxies, with an average increase of mass of between 15% (assuming equal 1:1 mergers) to 40% (assuming highly asymmetric mergers) in the densest D4 quartile, the importance of merging increases sharply with observed final mass, even for a merging rate that is independent of mass, simply because of the steepness of the mass-function. The history of galaxies is summarized for passive galaxies in Fig 16, where we show, as a function of their final mass, the fraction of galaxies initially quenched through different mechanisms, and whether or not they have subsequently merged.

## 7. DISCUSSION

The very simple purely empirical model that we have constructed in this paper was built upon a number of powerful "simplicities" observed in the galaxy population. We showed that it naturally produced several gross features of the galaxy population, and especially the Schechter form of the various galaxy mass-function(s), including the quantitative relationships between the various Schechter parameters. We find this success remarkable, given the very limited number of processes involved.

We stress that it is based on very few input "parameters": (1) the initial power-law slope α of the mass-function of star-forming galaxies, which is set in the model by the observed late epoch faint end slope $\alpha_s$ of blue galaxies; (2) the constant parameter μ in our star-formation rate quenching law, which is set by the observed M* of the galaxy population; (3) the slope β of the sSFR-mass relation, which is observed at both high and low redshift to be conveniently close to zero, and which must be close to zero if the faint-end slope of the star-forming mass-function changes little. At a greater level of detail, we can add (4) the small adjustment in the primordial mass-function in different environments that is required to get the correct relative normalization of the two $\alpha_s$ = – 0.4 and –1.4 components. To these four parameters are added the observed effects due to the development of structure in the Universe, which we have called "environment-" (or "satellite-") quenching, and merging. The parameters of both are set fairly directly by the observations and are not therefore "free" parameters.

The success of this very simple model in reproducing the observed mass-functions in SDSS is striking. While this empirical model does not offer new physical mechanisms for the main evolutionary changes, and cannot replace the more physically based SAM-type modeling or more physically realistic hydro-dynamical simulations, it nevertheless suggests which evolutionary signatures will be most rewarding to understand from a physical point of view.

Building on this success, we show, in this section, that our model also accounts at least qualitatively for several other characteristics of the population of passive galaxies, such as the mean age-mass relation for the passive galaxies, sometimes described as "anti-hierarchical", and the formation timescales of the stellar populations as revealed by the α-element abundances. We then explore two more general issues. But first we discuss the possible physical origin of mass-quenching.

*7.1 The ages and formation timescales of the stellar populations in passive galaxies.*

We define the "age" in this context to signify the V-band light-weighted age of the stellar populations (using the Bruzual & Charlot 2003 models to define the evolving M/L of a given stellar population). The more massive passive galaxies are produced by the mass-quenching star-formation term in the overall death function (equation 18). Their rate of production *at a given mass* therefore follows the average sSFR in the Universe and thus falls strongly at redshifts below $z \sim 2$. In contrast, the less massive passive galaxies with $m \ll$ M*, are primarily produced through the action of the mass-independent environmental quenching and merging processes, which together have a much weaker dependence on cosmic epoch. This therefore automatically produces a much broader distribution of ages, and a lower mean age. This is shown in Fig 17 for the simple model defined in Section 5.3.

The model would also predict that at a given stellar mass, passive galaxies that are satellites of other galaxies would be on average younger than passive galaxies that are "centrals". There is some evidence that this is indeed the case (e.g. Rogers et al., 2010). If our environmental-quenching effect is indeed associated with satellite-quenching, as we suspect, then it follows that central galaxies can *only* have been quenched through mass-quenching. Satellite galaxies on the other hand may *also* have been quenched through environment-quenching, i.e. satellite or merger quenching, which as shown in Fig 16, is a more recent phenomenon with a roughly flat rate over time.

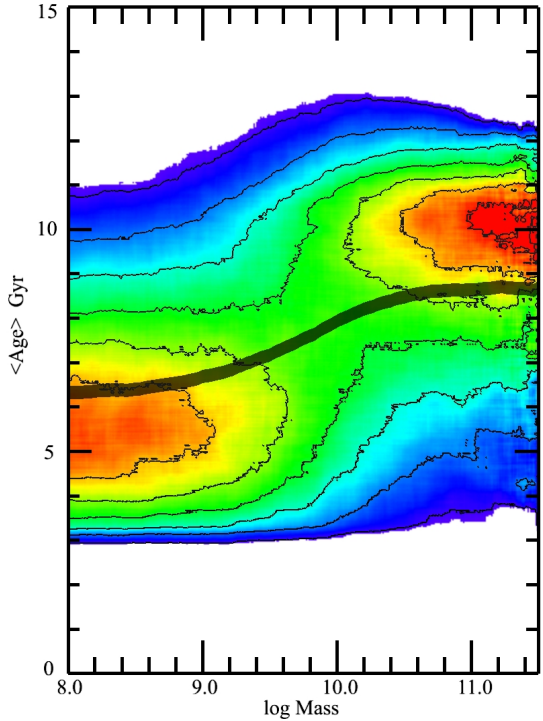 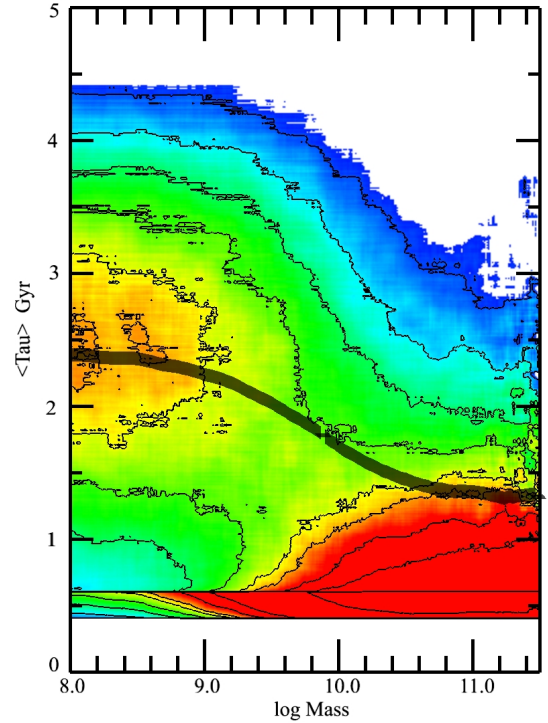

Fig 17: Distribution of the V-band light-weighted age of the stellar populations within passive galaxies at the present epoch, in the simple toy-model described in the text. This quantity is derived from the detailed star-formation history of each galaxy and gives the mean age of the constituent stellar populations. The solid black curve shows the median of the ages as a function of stellar mass of the final passive galaxy.

Fig 18: Distribution of the V-band light-weighted inverse sSFR of passive galaxies at the present epoch, in the simple toy-model. This quantity is derived from the detailed star-formation history of each galaxy and gives the typical timescale for enrichment of the stellar population of the galaxy. The solid black curve shows the median $\langle\tau\rangle$ as a function of mass.

Our quenching law (17) also naturally explains qualitatively the broad run of α-element abundances with the masses of passive galaxies (Thomas et al., 2005). The α-element abundance of a given set of stars in a galaxy will be related to the inverse sSFR of the galaxy at the time that those stars were formed, since this quantity sets the timescale for the chemical enrichment of those particular stars. In the simple picture sketched above, all passive galaxies that quenched prior to $z \sim 2$ will have formed *all* of their stars at high sSFR, specifically with $\tau_{sSFR} \sim 0.5$ Gyr. This is sufficiently short for them to be α-enhanced (Renzini 2009). Lower mass galaxies which form over a longer range of cosmic time (Fig 16) through our environment quenching effects, will have generally formed a significant fraction of their stars at later times (see Fig 15) when the sSFR in the galaxy population has fallen to lower values. The $\tau_{sSFR}$ will be longer, above 1 Gyr, and the galaxies should exhibit normal α/Fe ratios. In Fig 18, we show the distributions of the light weighted $\langle sSFR^{-1}\rangle$ for galaxies, i.e. averaging over the stars in a given galaxy, in the numerical model of Section 5.3 as a function of their final passive mass, again averaging over all environments.

These natural consequences of the death law given by equation (18) are in broad qualitative agreement with measurements of the ages and chemical compositions of passive galaxies (Carollo et al., 1993, 1994, Gallazzi et al. 2005, 2006; Thomas et al. 2005, 2010).

## 7.2 The role of the sSFR history as the cosmic clock

It will be clear from the foregoing that the precise cosmic sSFR history is more or less incidental to much of the discussion. Because the mass-quenching death rate is proportional to the star-formation rate, changes in the latter will be precisely balanced by the corresponding changes in the former. The resulting changes to the mass-function(s) will be driven by $\Delta \ln m$ (see equation 20), independent of the time-scale over which this occurs. If the death rate was indeed dominated by $\eta_m$ alone (i.e. without environmental effects) then a lower sSFR would simply cause a slower evolution, and not a different outcome as far as the build-up of a Schechter mass-function is concerned. The high and constant sSFR at early times ensures that the exponential growth of galaxies is sufficient to produce large numbers of star-forming galaxies around M* by the epoch corresponding to $z \sim 2$.

In practice, the juxtaposition of the star-formation driven $\eta_m$ (which depends strongly on mass at a given epoch) and the mass-independent $\eta_\rho$ terms in the composite η does introduce a dependence on the sSFR history. A change in the relative time-dependence of these two death rates would for instance change the relative size of the two components in the double Schechter function discussed in Section 5.2, even though the double form of the mass-function and the relationships between the M* and the $\alpha_s$ values would stay exactly the same.

One important consequence of the independence of our model on the precise history of the sSFR, is that the model is not dependent on the precise estimates of the star-formation rates in individual galaxies, which continue to have some measurement uncertainty. The basic form of the mass-quenching law is set by the observed constancy of M* of star-forming galaxies, while the numerical value of the constant μ is set by M*[-1]. Likewise, we could argue that the value of β *must* be close to zero simply *because* the mass-function of star-forming galaxies has almost constant $\alpha_s$ since $z \sim 2$. This conclusion is actually therefore independent of the actual measurements of SFR in individual galaxies and the precise form of sSFR($m,t,\rho$) is not so important.

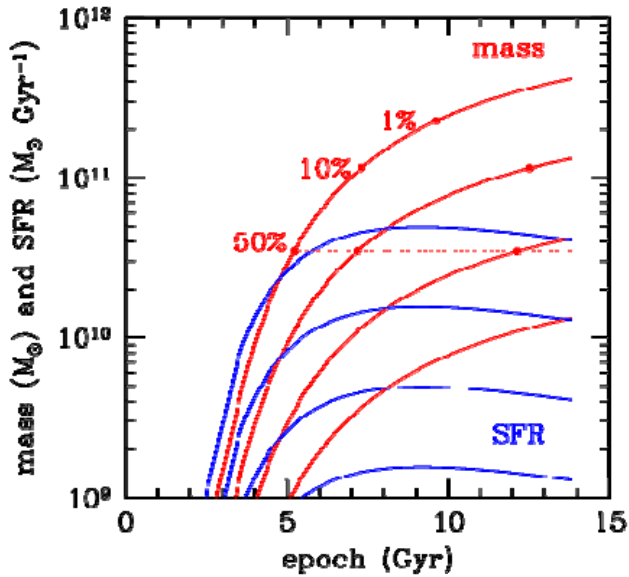

Fig 19: The growth in mass (red), and star-formation history (blue), of four galaxies of different initial mass that all follow the same global sSFR history given by equation (1). The steep continued increase of mass after z ~ 2 (3.5 Gyr) is evident. Although the sSFR steeply declines by a factor of 20 between z ~ 2 and the present, the SFR changes only a little for as long as the galaxies are not quenched. The dots along each mass curve show the points at which the survival probability P falls to 50%, 10% and 1%. The masses of all four galaxies is the same at each of these probability thresholds, even though they occur at very different epochs. In this view, the Schechter function arises as the result of the "universal" P(m).

*7.3 Is our mass-quenching scenario simply a reflection of a limiting mass for galaxies or haloes?*

We have tried to stress throughout the paper the empirical nature of our analysis, without attempting to ascribe strong causal connections to particular physical processes. One of the great successes of our model is clearly the natural emergence of the various Schechter functions that describe star-forming and passive galaxies. In this section we explore the links to other possible mechanisms, and in particular the idea of a "limiting maximum mass" for galaxies. Such a limit could be related for example to the well-established idea of cooling of gas in dark matter haloes as a function of mass (Silk 1976, Rees & Ostriker 1977).

The role of equation (23) in producing the Schechter function exponential cut-off can be viewed in terms of survival probabilities. We introduced η as a probabilistic mass-quenching rate for a given galaxy at a given time. If we consider some galaxy that is growing in mass via star-formation, then the probability $P(m)$ that it will survive to a particular stellar mass $m$, will simply be given by:

$$\frac{dP}{dt} = -\eta\, P = -\mu \frac{dm}{dt} P$$
$$\frac{dP}{P} = -\mu\, dm$$
$$P \propto \exp(-\mu m) = \exp(-m/M^*) \qquad (29)$$

It can be seen that equation (27) naturally gives the Schechter cut-off to the mass-function of galaxies at $M^* = \mu^{-1}$ in terms of the survival probability distribution function of individual galaxies. It should be noted how $P(m)$ is completely independent of the form or history of $dm/dt$. This is also why our model does not depend on the detailed cosmic evolution of the sSFR. It also ensures that the model is independent of the scatter around the sSFR-mass relation and of the observational estimates of star-formation rates in distant galaxies.

This independence of the outcome on the details of *individual* star-formation histories also means that we do not need to consider explicitly any extreme "star-bursting" galaxies, such as the ULIRGs or sub-millimeter galaxies, which will have individual sSFR above the "main-sequence" relation. Provided that they are quenched according to the same quenching law that is given by equation (17) they will automatically be included in the framework. This is the justification for not considering their role(s) earlier in the paper.

Exploring this further, it is interesting to look at the mass and SFR of a particular galaxy that follows the changing sSFR described by equation (1) (see also Renzini 2009). This is shown in Fig 19. During the period that the sSFR is constant, at $z > 2$, the galaxy will increase its mass exponentially. However, as the sSFR starts to decline, the rate of mass growth slows, but the SFR, and thus the quenching probability, for this individual galaxy stays broadly similar over an extended period of time. This is because the build-up of stellar mass largely compensates for the decline in sSFR. Because of the balancing of mass and sSFR, one could therefore view quenching as a quasi-random process after $z \sim 2$. A more massive galaxy, following a higher track in Fig 19 will also experience a more or less constant risk of being quenched per unit time, after $z \sim 2$, but at a higher level. A lower mass galaxy will continue to grow without serious risk of quenching until its mass approaches M*. In the end the total mass of stars that is produced (statistically) by the time all three galaxies are finally quenched, will be the same, and be roughly equal to M* as in equation (25). This is again a consequence of the survival probability analysis given in equation (25).

It might therefore be thought that there is something of a tautology in our quenching law. A clear consequence of it is that the statistical lifetimes of individual galaxies are simply set by the time that is needed to build up an M*-worth of stellar mass, regardless of how fast or slowly this is done. Since most galaxies will be the dominant galaxies in their haloes, and since we are anyway in a sense taking care of the satellites through our "environment-quenching" term, we could also broadly translate this to simply requiring quenching when the mass of the *dark matter halo* reaches a certain limiting value. This limit could be implied by well-established cooling time arguments (Silk 1976, Rees & Ostriker 1977, Dekel & Birnboim 2006, Cattaneo et al. 2006).

It is clear from the above that our empirical "star formation rate" quenching law and a "mass-threshold" quenching law are rather similar in their outcomes, even if they are presumably based on very different physics. Both involve a rather impressive constancy since $z \sim 2$: SFR-quenching requires that the physical relationship of quenching rate to star-formation rate be independent of epoch (as well as the mass and environment of a galaxy), while the "threshold-quenching" requires that the mass-threshold be likewise rather precisely constant over a very wide range of time and environments, although this is apparently achieved in some models (e.g. Birnboim & Dekel, 2003).

The only differences might be in the detailed outcomes. We refer back to the point made earlier in the paper. Our SFR-quenching law, because of its implicit mass-dependence (roughly proportional at any epoch to mass, with β ~ 0) and its implicit sSFR linkage, naturally produces the precisely Schechter forms of the mass-function of passive and active galaxies. This is true both for the exponential cut-off *and* for the change in faint end power-law slope with $\Delta\alpha_s \sim 1$ of the passive galaxies relative to the star-forming galaxies. The Schechter forms for the mass-function(s) are seen over about two orders of magnitude in galactic mass.

If the build-up of passive galaxies was simply related to some threshold in mass beyond which galaxies cannot grow, then we might well have expected some other form for the mass functions of the resulting galaxies. An absolute "brick-wall" for galaxy growth would have resulted in a step-function for the blue galaxies and a delta function for the passive ones. More realistically, we might have expected a Gaussian distribution for the mass-function of passive galaxies, reflecting the statistical variations in the limiting mass due to different random factors. The fact that we get an exact Schechter functions for both the star-forming *and* for the mass-quenched passive galaxies, which clearly holds in the SDSS mass-function (Fig 12) over about 2 orders of magnitude in stellar mass, is a natural consequence of a probabilistic death rate that operates over a very broad range of masses in a way that is, at any epoch, closely proportional to the product of mass and specific star formation rate, i.e. to the star formation rate alone.

It is not clear that a "mass-limit" quenching law can so successfully account for the Schechter shapes of the different mass-functions and the precise inter-relationships shown above. Not least, we emphasize that the basic separability between mass and environment, that was established in this paper, requires that any "second-parameter" producing a scatter in the dark halo mass-limit must be strictly independent of the Mpc-scale environment.

## 8. SUMMARY AND CONCLUSIONS

We have studied and compared the global relationships between star-formation rate, stellar mass and Mpc-scale environment in SDSS and zCOSMOS, searching for simple representations of the data that lead to correspondingly simple empirical descriptions of the dominant evolutionary changes in the galaxy population. We then explore the consequences of these empirical facts.

We first emphasize the tight dependence of star-formation rate on stellar mass that has been seen in the SDSS and in the high redshift galaxy population to $z \sim 2$. This is characterized by a specific star-formation rate (sSFR) that is only weakly dependent on stellar mass, does not appear to depend on environment, but which does evolve strongly with epoch. Our main conclusions are simplified by the observed environment independence and (at least approximate) mass independence of the sSFR, but are largely independent of its precise temporal evolution or of the scatter around the SFR-*m* relation.

The main focus of the paper is then the quenching transitions between star-forming and passive galaxies. Powerful new insights have come from a new formalism that looks at the differential effects of mass and Mpc-scale environment on the fraction of galaxies that are quenched and from consideration of the mass-function of star-forming galaxies.

Our results and conclusions may be summarized as follows.

1. In SDSS we demonstrate the clear separability of the *differential* effects of stellar mass and environment on the fraction of galaxies that are actively forming stars compared with those which are passive. The differential effects of the environment do not depend on the mass of the galaxies and, vice versa, the differential effects of mass do not depend on the environment. This suggests two different effects may be operating, which we refer to as "mass quenching" and "environment quenching".

2. This separability between the effects of mass and environment is also seen in the zCOSMOS sample out to $z = 1$. Remarkably, the differential effects of the environment, at fixed over-density δ, are found to be independent of epoch to $z \sim 1$. The emergence of the environmental dependence of $f_{red}$ as time passes is due to the migration of galaxies to higher over-densities as large scale structure grows in the Universe, and not to any temporal change in the environment effect at fixed over-density δ.

3. This suggests that the environment acts through a "once-only" process as the environment of a given galaxy changes. A natural possibility is "satellite-quenching" as galaxies fall in to larger haloes, since the satellite/central fraction in mock catalogues is, like $\varepsilon_\rho$, strongly dependent on δ but independent of epoch (at least at $z < 1$), and galactic mass (for masses below $10^{10.9}$ $M_\odot$). There is indeed a good correspondence between our $\varepsilon_\rho(\rho)$ parameter and $f_{sat}(\rho)$, implying that about 30-70% of satellites are quenched through satellite-quenching over the observed range of environments.

4. In contrast, the mass-quenching process must be continuously operating and be governed by a probabilistic transformation rate. We stress the empirical fact that the mass-function of star-forming galaxies is a Schechter function, with essentially constant M* since $z \sim 2$, an almost constant faint end slope $\alpha_s$, and a value of $\phi^*$ that only slowly increases with time. This demands that the rate at which galaxies die through "mass quenching" must be proportional to their individual star formation rates,

$$\eta_m = \mu\ SFR$$

at least for galaxies around and above M*. This empirical law could conceivably be the result of different physical processes, un-connected with star-formation, which nevertheless combine to mimic this very simple form. Regardless, this simple description must hold over a wide range of epochs ($0 < z < 2$) despite the large decline in specific star-formation rates (by a factor of about 20), and the increase in the characteristic masses of dark matter haloes (by two orders of magnitude) during this time interval.

5. Such a quenching law naturally establishes and maintains a Schechter-function for star-forming galaxies, with a characteristic mass M* that is set by

$$M^* = \mu^{-1}.$$

We suggest that this mechanism(s) provides the explanation of the exponential cut-off in the Schechter function and sets the (constant) value of M*. Having established this law in the high mass regime for *star-forming* galaxies, we postulate that it actually applies also at lower masses. The evidence in favor of this come from the Schechter form of the mass-function of *passive* galaxies.

6. The physical mechanism which causes $\eta_m$ to be given by the SFR is not known. If there is indeed a direct causal link to star-formation, it could reflect a feedback mechanism, linked either directly to star-formation, or to AGN, since the rate of black-hole accretion in galaxies appears to be proportional to the SFR in galaxies, at least in a statistical sense. Mass quenching dominates the quenching of massive passive galaxies $m > 10^{10.2}$ $M_\odot$.

7. Our empirical mass-quenching law will produce a Schechter function for passive galaxies which will have exactly the same M* as that of the star-forming galaxies, but a faint end slope that is shallower (less negative) by an amount related to the slope of the sSFR-mass relation, β, i.e. $\Delta\alpha = (1+\beta) \sim 0.9$, or $\Delta\alpha = 1.0$ if β = 0. In other words the difference in faint end slope directly reflects the mass dependence of the star formation rate at a given epoch, which is also responsible for the establishment of the exponential cut-off at higher masses.

8. There will also be a (mass-independent) quenching rate associated with the gradual development of the environmental quenching effects, i.e. satellite quenching, plus the (assumed mass independent) merging of galaxies. Merging evidently dominates at early times, and satellite quenching at late times (presumably as intra-halo densities drop), but both of course represent different outcomes of the same dark matter halo assembly. Together, they dominate the quenching of star-forming galaxies at low masses, $m < 10^{10}$ M$_\odot$.

9. The mass independent death rate associated with these general environmental effects produces a second Schechter component for passive galaxies, with the same $\alpha_s$ and M* as the star-forming galaxies. In combination, we therefore expect a double (two component) Schechter mass-function for passive galaxies, both with exactly the same M* as the star-forming galaxies. The two components reflect the two quenching routes, one that is independent of mass and the other that is (at a fixed epoch) roughly proportional to mass.

10. Furthermore, the combination of the single Schechter function of star-forming galaxies with the double Schechter mass-function for passive galaxies, all with the same M*, and two components with the same faint end slope, also therefore naturally produces the two-component double Schechter mass-function for galaxies as a whole.

11. Subsequent post-quenching dry merging of galaxies will slightly modify the expected mass functions in the higher density regions where such merging will be more important. Assuming this is again mass-independent, the effect on the mass-function of low mass passive galaxies (produced by satellite and merger quenching) will be unnoticeable. For the dominant population of massive galaxies around M* that are produced by mass-quenching, the effect will be to cause a migration of M* towards higher values and of $\alpha_s$ to slightly more negative values.

12. All of the detailed inter-relationships of the Schechter parameters that are predicted by our simple model for the blue and red galaxies in different environments are seen to impressive precision in our analysis of the mass-functions in the SDSS. Specifically, we see
   (a) the same M* for red and blue galaxies, and a change in the faint end slope $\Delta\alpha_s \sim 1$, in the low density D1 quartile, as predicted;
   (b) the same M* and same $\alpha_s$ for blue galaxies in the D1 and D4 quartiles, as predicted;
   (c) small changes in the M* and $\alpha_s$ for the dominant population of passive galaxies in D4 relative to D1, that are consistent with the effects of a small amount ($\Delta m \sim 15\%$ and definitely less than 40%) of subsequent dry-merging in D4.
   (d) the increase with environmental density in the amplitude of the second Schechter component of passive galaxies, as expected.

13. Our simple empirical model, which has only three or four free-parameters, is therefore able to reproduce the mass function(s) of the different components of the galaxy population in different environments, and the $f_{red}$ as functions of mass and environment. While it is by design purely empirical, it suggests the clear observational signatures of the most important processes driving galaxy evolution that must be reproduced by more physically based models.

14. The model makes a clear prediction for the mass-function of any transient objects that are seen during the process of being mass-quenched. Their mass-function should have the same shape as that of the existing passive galaxies, but a number density that is the product of the $\phi*$ of the continuing star-forming galaxies multiplied by the dimensionless ratio of the visibility timescale of the transient phenomenon and the characteristic star-formation timescale, evaluated at M*.

15. Our empirical model also naturally explains why the mean age of passive galaxies increases "anti-hierarchically" with stellar mass, and the why the spread of stellar ages within passive galaxies, and the degree of α-element enrichment, both decrease with increasing stellar mass, as observed in the local galaxy population. The model also predicts that the passive galaxies which are satellites should be younger than those which are not.

16. Finally, we have noted the links to models in which the build-up in stellar mass in galaxies is limited at some threshold mass-scale, perhaps associated with cooling of gas in haloes of different masses. We argue however that the scheme outlined above may more naturally account for the precise Schechter form of the mass-function(s) that is seen over two orders of magnitude in the mass-function.

We stress that the above conclusions, while undoubtedly ignoring much of the detailed processes of galaxy evolution, follow rather directly from the observed properties of the galaxy population that we have identified. Furthermore, although it is empirically defined only for $z < 2$, and in some important respects only for $z < 1$, we find that this simple picture appears to work very well at higher redshifts also.

We therefore conclude that there are, from an empirical stand-point, four main drivers of galaxy evolution operating over a broad span of cosmic time:

(1) a poorly understood physical process that sets the uniform sSFR across the whole galaxy population and also presumably controls its evolution with redshift; this effectively sets the "cosmic clock" for the evolution of the galaxy population but does not have a big influence on the eventual outcome. This is probably linked to the overall gas supply and the accretion of baryons (and dark matter) onto growing haloes.

(2) an unknown physical process, or set of processes, possibly involving feedback of some sort, that "mass-quenches" galaxies at a rate that has, or at least mimics, a simple proportionality to galaxies' individual star-formation rates; this naturally produces the Schechter form of the mass-function of star-forming galaxies and sets the numerical value of M*, and establishes the dominant population of passive galaxies on the red sequence, setting up the same M* but with a distinctively modified faint-end slope.

(3) the hierarchical assembly of dark matter haloes, which modifies the galaxy population principally at lower masses through, initially, the merging of galaxies and, subsequently, through the "environment-quenching" of galaxies that do not merge during the assembly of dark matter haloes; these two processes produce a second Schechter function of passive galaxies, and the progressive appearance of environmental differentiation in the galaxy population, especially at lower masses and later epochs. In combination with the mass quenching, this process also explains a number of other properties of passive galaxies, including the ages and formation timescales of their constituent stellar populations.

(4) the relatively minor effects (at least on the mass-function) of subsequent post-quenching "dry-merging", which increases the mass of passive galaxies in the denser environments by a modest amount, on average by between 15% and 40% in the highest density quartile.

## ACKNOWLEDGEMENTS

The zCOSMOS survey was undertaken at the ESO VLT as Large Program 175.A-0839. We gratefully acknowledge the work of many individuals, not appearing as authors of this paper, whose work has enabled large surveys such as COSMOS and the SDSS. We also thank Sebastiano Cantalupo for a very helpful and critical reading of an earlier draft of this paper, and the anonymous referee for a helpful and sympathetic reading. We gratefully acknowledge NASA's IDL Astronomy Users Library, the IDL code base maintained by D. Schlegel, the *kcorrect* package of M. Blanton, and the star formation rates from J. Brinchmann taken from the MPA website. This work was supported in part by the Swiss National Science Foundation.